\documentclass[twocolumn, nopreprintnumbers, 
               superscriptaddress,
               eqsecnum,nofootinbib]{revtex4}
\usepackage{graphicx, fancybox,ulem}
\usepackage{amsmath,amssymb}
\usepackage[ colorlinks=true, pdfstartview=FitV, linkcolor=red, citecolor=blue, urlcolor=blue]{hyperref} 
\usepackage{color} 
\usepackage{bm}

\newcommand{\vect}[1]{\mathbf{#1}}

\newcommand{\im}{\mathrm{Im}}

\newcommand{\vp}{\vect{p}}

\newcommand{\vk}{\vect{k}}

\newcommand{\vx}{\vect{x}}

\newcommand{\vzero}{\vect{0}}

\newcommand{\muf}{\mu_f}
\newcommand{\mub}{\mu_b}
\newcommand{\Nf}{N_f}

\newcommand{\nf}{n_f}

\newcommand{\rhof}{\rho_f}
\newcommand{\rhob}{\rho_b}

\newcommand{\ef}{\epsilon^f_\vk}

\newcommand{\kcf}{k_{cf}}
\newcommand{\kf}{k_{F}}
\newcommand{\eF}{\epsilon_F}
\newcommand{\vol}{\Omega}

\newcommand{\GRPA}{G^{\mbox{\tiny RPA}}}
\newcommand{\GMFpole}{G^{\mbox{\tiny MF}}_{\rm pole}}
\newcommand{\GMFcont}{G^{\mbox{\tiny MF}}_{\rm cont}}

\newcommand \beq{\begin{eqnarray}}
\newcommand \eeq{\end{eqnarray}} 
\newcommand{\nn}{\nonumber\\ }
\def\bra#1{\langle#1\vert}
\def\ket#1{\vert#1\rangle}
\newcommand{\rmd}{{\rm d}}

\newcommand{\comment}[1]{}
\def\comment#1{\color{red}{\bf{#1}}}

\begin{document}
\title{Goldstino in supersymmetric Bose-Fermi mixtures in the presence of Bose Einstein condensate}

\author{Jean-Paul Blaizot}
\affiliation{Institut de Physique Th\'eorique, CEA/Saclay,  CNRS/UMR 3681,  F-91191 Gif-sur-Yvette Cedex, France}

\author{Yoshimasa Hidaka}
\affiliation{Theoretical Research Division, Nishina Center,
RIKEN, Wako 351-0198, Japan}

\author{Daisuke Satow}
\email{dsato@th.physik.uni-frankfurt.de}
\affiliation{Goethe University Frankfurt am Main, Institute for Theoretical Physics, Max-von-Laue-Str. 1, D-60438 Frankfurt am Main, Germany}

\begin{abstract} 
We analyze the spectral properties of the Goldstino excitation in a Bose-Fermi mixture of cold atoms, whose masses and interaction strengths are tuned so that the hamiltonian is supersymmetric. We consider systems at zero temperature and assume that, in the weak coupling regime, the fermions form a Fermi sea, while the bosons  form a Bose-Einstein condensate. We study the excitation spectrum within a simple extension of the random phase approximation, taking into account the mixing between the supercharge and the fermion caused by the condensate. This mixing affects the fermion spectrum strongly. We argue that the corresponding modification of the fermion spectrum, and the associated fermion distribution in momentum space, could be accessible experimentally, and potentially allow for a determination of the Goldstino properties. 
\end{abstract} 

\date{\today}


\maketitle

\section{Introduction}
\label{sec:intro}

The possibility to prepare low temperature Bose-Fermi mixtures with tunable interactions~\cite{Ferrier-Barbut, Hara} is offering new playgrounds for the study of novel phenomena in many-body systems. 
Of interest to us in the present paper are mixed systems of bosons and fermions  exhibiting (a restricted form of) supersymmetry~\cite{Wess:1974tw}, that is, an invariance under the interchange of fermions and bosons~\cite{Yu:2010zv}. In such systems, and when the supersymmetry is explicitly broken~\cite{Das:1978rx}\footnote{
Even in systems that have only an approximate supersymmetry, such as the Yukawa model, quantum electrodynamics, and the quantum chromodynamics at ultrarelativistic temperature, the existence of a quasi-Goldstino was suggested and its properties studied in~\cite{Hidaka:2011rz, Satow:2013oya, Blaizot:2014hka}.}, which occurs for instance when the chemical potentials for the bosons and the fermions are different, one expects a new type of long wavelength collective excitation carrying fermionic quantum number. This excitation, which shares many properties with the more familiar Nambu-Goldstone boson~\cite{Nambu:1961tp, Goldstone:1961eq}, has  been dubbed a Goldstino~\cite{Kratzert:2003cr}\footnote{In fact, the existence of a Goldstone fermion associated with the spontaneous breaking of supersymmetry was considered already in the early days of supersymmetry~\cite{ Fayet:1974jb}. }
The possible realization of the  Goldstino in cold atom systems was suggested in Ref.~\cite{Yu:2007xb} and the spectral properties of the Goldstino have been analyzed in Refs.~\cite{Shi:2009ak, Blaizot:2015wba, Lai:2015fia, Sannomiya:2016wlz}.

The special influence of a Bose-Einstein condensate (BEC) on the spectral properties of the Goldstino has not been thoroughly discussed so far, although it has been suggested that the presence of a BEC could lead to an easier  experimental detection of the Goldstino~\cite{Yu:2007xb}:
This is because the operator ($q$) that excites the Goldstino is essentially composed of the boson creation operator ($b^\dagger$) and the fermion annihilation operator ($f$), $q\simeq b^\dagger f$.
In the BEC phase, $b$ is dominated by the condensate part, which is a $c$-number. It follows therefore that the Goldstino operator contains a term  proportional to $f$. This suggests that the spectral properties of the Goldstino are reflected in those of the fermion, and 
the latter can be observed in photoemission spectroscopy~\cite{Stewart: 2008}. This provides motivation for further theoretical investigation of the spectral properties of the Goldstino in the presence of a BEC. This is the purpose of the present paper, which extends our previous work, limited to two dimensions, and where therefore BEC was absent~\cite{Blaizot:2015wba}.

This paper is organized as follows:
In the next section, we briefly introduce a simple model for a Bose-Fermi mixture of cold atoms. The model parameters can  be  tuned so as to achieve supersymmetry, in which case a Goldstino excitation emerges in the spectrum. 
In Sec.~\ref{sec:BEC}, we discuss how the interaction terms in the Hamiltonian are organized in order to treat the BEC. Conditions on the physically acceptable range of parameters are also discussed. 
In Sec.~\ref{sec:spectrum}, we analyze the  components of the Goldstino spectral function, paying particular attention to the mixing  between the supercharge and the fermion. This is done at weak coupling using a simple extension of the random phase approximation (RPA), for zero and finite momentum. 
In Sec.~\ref{sec:phenomenology}, we argue on  the modification of the fermion spectrum, and how it is reflected in the fermion distribution function.
Measurement of these quantities could yield information on the Goldstino properties. 
The last section contains a brief summary of the paper.
In the Appendix, we show that the contribution from the phonon to the modification to the Goldstino spectrum is negligible compared with that from the boson.

In this paper, we use units with $\hbar=k_B=1$. 

\section{A simple model for the Goldstino}
\label{sec:intro2}

In this section, we  introduce the simple model for a Bose-Fermi mixture  on which our discussion will be based. 
The hamiltonian of this model has the generic form 
\beq
H=H_{f}+H_{b}+V,
\eeq 
where 
\begin{align}
H_{f}&= \frac{1}{2m_f}\int d^3\vx \left(\nabla f^\dagger(\vx)\right) \nabla f(\vx),\\ 
H_{b}&= \frac{1}{2m_b}\int d^3\vx \left(\nabla b^\dagger(\vx)\right)\nabla b(\vx) ,\\
V&= \int d^3 \vx \left[\frac{U_{bb}}{2}b^\dagger(\vx)b^\dagger(\vx)b(\vx)b(\vx)
+U_{bf} n_b(\vx) n_f(\vx) \right],
\end{align}
with $n_f(\vx)=f^\dagger(\vx) f(\vx)$ and $n_b(\vx)=b^\dagger(\vx) b(\vx)$ are the densities of the fermions and the bosons. This hamiltonian  can be viewed  for instance as the long wavelength limit of the lattice hamiltonian used in Ref.~\cite{Yu:2007xb}. 
Note that there is no interaction among the fermions: this is because we assume  the fermion spin to  be  polarized so that there is only one active spin degree of freedom.
By adding the chemical potential terms, we obtain the grand canonical Hamiltonian, 
\begin{align}
H_G
&\equiv H-\mu_f N_f-\mu_b N_b
=H-\mu N-\Delta\mu\Delta N,
\end{align}
where $N_f\equiv \int d^3\vx ~n_f(\vx)$, $N_b\equiv \int d^3\vx ~n_b(\vx)$, $N\equiv N_f+N_b$, $\Delta N\equiv (N_f-N_b)/2$, $\mu\equiv (\mu_f+\mu_b)/2$, and $\Delta\mu\equiv \mu_f-\mu_b$.

The supercharge operator is defined as\footnote{Note that we interchange here  the definitions of  $q$ and $q^\dagger$ that we used in our  previous work~\cite{Blaizot:2015wba}: in the present definition,  $q$ creates a boson instead of a fermion.
In the presence of BEC, this convention turns out to be more convenient. In particular it makes the fermion propagator appears naturally in the decomposition of the Goldstino propagator, see Eq.~(\ref{eq:GS-SR}) below.} 
\beq
Q&\equiv \int d^3\vx ~q(\vx),~~~
q(\vx)\equiv f(\vx) b^\dagger(\vx).
\eeq
The operator $q(\vx)$ replaces locally a fermion by a boson.
It is easy to show that the grand canonical hamiltonian is supersymmetric when $m_f=m_b=m$ and $U_{bf}=U_{bb}=U$, except for the chemical potential difference term: $[H_G,Q]=\Delta\mu Q$.
In the rest of the paper, we will focus on this particular case. 
For the convenience of the reader, we also translate $U$ into the scattering strength, $a$:
By introducing $a_{bb}\equiv U_{bb} m_b/(4\pi)$ and $a_{bf}\equiv U_{bf} m_{bf}/(4\pi)$, where $m_{bf}\equiv 2m_f m_b/(m_b+m_f)$, the condition of supersymmetry can be written as $m_{bf}=m$ and $a_{bb}=a_{bf}=a$.
Note that the action of the supercharge density $q^\dagger$ on a state with a given number of fermions and bosons leaves the total number of atoms unchanged, but increases $\Delta N$ by one unit. Similarly the action of $q$ decreases $\Delta N$ by one unit.

In order to study the excitations induced by the supercharge, we shall focus on the retarded Green's function,
 \begin{align} 
\label{eq:definition-retarded}
\begin{split}
G^R(x)&\equiv i\theta(t)\langle \{ q(t,\vx), q^\dagger(0)\}\rangle,
\end{split}
\end{align} 
where the angular brakets denote an average over the ground state of the system. 
Its Fourier transform is written as
\begin{align} 
\label{eq:definition-retarded2}
\begin{split}
G^R(p)&=  i\int dt\int d^3\vx\, e^{i\omega t-i\vp\cdot\vx}
\theta(t)\langle \{ q(t,\vx), q^\dagger(0)\}\rangle.
\end{split}
\end{align}
Here we have introduced a 4-vector notation, to be used throughout: $x^\mu\equiv (t,\vx)$ and $p^\mu\equiv (\omega,\vp)$. The frequency $\omega$ is assumed to contain a small positive imaginary part $\epsilon$ ($\omega\to \omega+i\epsilon$) in order to take into account the retarded condition. Such a small imaginary part will not be indicated explicitly in order to simplify the formulae.  In fact, we shall also most of the time drop the superscript $R$, and indicate it only when necessary to avoid confusion.

Let us recall some general features of this Green's function by looking at its spectral representation in terms of the excited states $\psi_n$ and $\psi_m$ that can be reached from the ground state by acting respectively with $q^\dagger(\vp)$ and $q(\vp)$, where 
\beq\label{superchargep}
q(\vp)=\sum_\vk f_\vk b^\dagger_{\vk-\vp}.
\eeq
We obtain from Eq.~(\ref{eq:definition-retarded2})
\beq
\label{eq:GR-qq*}
&&G(\omega,\vp) =-\frac{1}{\Omega}\left\{ \sum_n \frac{ \bra{\psi_0}q(\vp)\ket{\psi_n}\bra{\psi_n}q^\dagger(\vp)  \ket{\psi_0}}{\omega-(E_n-E_0-\Delta\mu)} \right.\nn
&&\qquad\qquad \qquad\quad\left. +\sum_m \frac{ \bra{\psi_0}q^\dagger(\vp)\ket{\psi_m}\bra{\psi_m}q (\vp) \ket{\psi_0} }{\omega+(E_m-E_0+\Delta\mu )}  \right\}, \nn
\eeq
where $\Omega$ is the volume of the system.
This expression shows that $G(\omega,\vp)$ has poles at $\omega=E_n-E_0-\Delta\mu$ corresponding to the free energies of the states $\ket{\psi_n}$ that have non vanishing overlap with $q^\dagger(\vp)  \ket{\psi_0}$, and at $-\omega=E_m-E_0+\Delta\mu$  corresponding to the free energies of the states $\ket{\psi_m}$ that overlap with $ q(\vp)  \ket{\psi_0}$. Stability (in Fock space) requires that the two sets of poles sit respectively at positive or negative values of $\omega$. 
Note that $\omega$ always appears as $\bar\omega\equiv \omega+\Delta\mu$ in $G(\omega,\vp)$.

In the supersymmetric case, it can be  shown that $G(\omega,\vp=\vzero)$  has the following form~\cite{Blaizot:2015wba}:
\beq
\label{eq:G0-p=0-pole}
 G(\omega,\vp=\vzero)= -\frac{\rho}{\omega+\Delta\mu}=-\frac{\rho}{\bar\omega}
\eeq
where we have set $\rho\equiv \langle n(\vx)\rangle=\langle n_f(\vx)+n_b(\vx)\rangle$, the equilibrium density $\langle n(\vx)\rangle$ being assumed uniform, i.e., independent of $\vx$.  
Since the density can be expressed as $\rho=\langle\{Q,q^{\dag}(\vx)\}\rangle=\langle\{Q^{\dag},q(\vx)\}\rangle$,  $\rho$ plays the role of the order parameter associated with the spontaneous breaking of supersymmetry. Note that although both $Q$ and $Q^{\dag}$ are broken charges, this does not mean that there appears two independent Goldstinos. 
Nonvanishing expectation value of anti-commutator between charges, $\langle \{Q,Q^{\dag}\}\rangle/\Omega\neq0$, implies indeed that $Q$ and $Q^{\dag}$ are canonically conjugate~\cite{Nambu:2004yia}, and each charge does not generate an independent Goldstino. Such a mode is referred to a type-B mode~\cite{NG}\footnote{We note that unlike Nambu-Goldstone modes, a type-B Goldstino does not necessarily have a quadratic dispersion relation~\cite{Sannomiya:2016wlz},  even though we shall find out that our Goldstino has a quadratic one.}. 
 The expression~\eqref{eq:G0-p=0-pole} reveals  that the retarded propagator has a single pole at $\omega=-\Delta\mu$, i.e. at $\bar\omega=0$, with $\Delta\mu$ being the source of the explicit breaking of supersymmetry. 
 This is the Goldstino pole.
 The existence of this pole follows directly from the conservation law and the canonical (anti-) commutation relations~\cite{Yu:2007xb, Blaizot:2015wba}. It 
  does not depend on the details of the Hamiltonian. 
This is the Goldstino's counterpart to the gapped Nambu-Goldstone modes~\cite{Nicolis:2012vf,Nicolis:2013sga,Watanabe:2013uya}. 
  In the following, we shall often refer to $G(\omega,\vp)$ as the Goldstino propagator. 
 
We shall also be interested in the associated Goldstino spectral function 
\beq
\sigma(\omega,\vp)=2\im \,G(\omega,\vp) .
\eeq
This  spectral function obeys simple sum rules~\cite{Blaizot:2015wba}.
The first sum rule determines the zeroth moment of the spectral function.
It is valid regardless of the details of the Hamiltonian, and 
reads
\begin{align}
\label{eq:sumrule1-G}
\int \frac{d\bar{\omega}}{2\pi}\sigma(p)
&= \rho.
\end{align}
Another sum rule (analog to the``$f$-sum rule'') gives the first moment of the spectral function
\begin{align}
\label{eq:sumrule2-G}
\int \frac{d\bar{\omega}}{2\pi}\bar{\omega} \sigma(p)
&= \alpha_s \frac{\vp^2}{2m} \rho,\qquad \alpha_s\equiv \frac{\rhob-\rhof}{\rho}.
\end{align}
Note that the right-hand side is independent of the interaction strength $U$.

If we were to assume that the spectral function is dominated by a single peak, the sum rules (\ref{eq:sumrule1-G}) and (\ref{eq:sumrule2-G}) would give us 
\begin{align}
 \sigma(p)
 &= 2\pi \rho\, \delta \left(\bar{\omega}-\alpha_s \frac{\vp^2}{2m}\right) .
\end{align}
In this case, $\alpha_s$ would completely determine the dispersion relation of the Goldstino~\cite{Bradlyn:2015kca}.
However, we will see that this assumption is not valid, at least in the weak coupling case (see Sec.~\ref{sec:spectrum}).
\section{BEC}
\label{sec:BEC}

In order to progress further, we need to specify the ground state of the system through which the Goldstino propagates. We shall assume in this paper that this ground state is, in the absence of interactions, the product of a Fermi sea of fermions and a coherent state of bosons, with the boson occupying the zero momentum state. We shall then proceed to the analysis of the effects of the interactions, assuming that the coupling is small. As we shall see, we need to go beyond strict perturbation theory, and implement various resummations in order to account properly for the relevant processes. 

The calculations to be presented in the next sections, depend on a number of parameters. 
Because of the assumed supersymmetry, the Hamiltonian itself depends on two parameters, the mass $m$ of the atoms, the same for the bosons and the fermions, and the coupling strength $U$. 
The system, in addition, depends on the densities of the fermions and the bosons, respectively $\rho_f$ and $\rho_b$. 
We shall focus on cases where the densities are of comparable orders of magnitude. 
Thus, in all the calculations, we have chosen $\rho_b=2\rho_f$, as in our previous work \cite{Blaizot:2015wba}. 
The Fermi energy, $\varepsilon_F=k_F^2/(2m)$ provides a convenient unit for the energies. 
Here $k_F$ is the Fermi momentum related to the fermion density via $\rho_f=k_F^3/(6\pi^2)$. 
The quantity $U\rho$ has the dimension of an energy, and the ratio $U\rho/\varepsilon_F$ can be used as a measure of the strength of the coupling, with weak coupling implying $U\rho\ll \varepsilon_F$.


In order to treat the condensate in the weak coupling approximation, it is convenient  to isolate the operator $b_0 \equiv b_{\vp=\vzero}$ from the finite momentum operators $b_\vp$. Recall that the expectation value of $b_0$ in the Bose condensate is non vanishing, $\langle b_0\rangle =\sqrt{N_0}$, with $N_0$ the number of bosons in the condensate. One may then write
\beq
\label{eq:BEC-boson-decompose}
b_{\vp=\vzero} =\frac{1}{\sqrt{\Omega}}\int \rmd^3\vx \, b(\vx)\rightarrow \sqrt{N_0}+\tilde b_{\vp=\vzero},
\eeq
where $\tilde b_{\vp=\vzero}$ represents the fluctuation part of the operator. 
For the approximations that we shall use later, it can be neglected. Similarly,  the depletion of the condensate due to the interactions will be ignored in leading order, so that the boson density is  simply  $\rhob= N_0/\Omega$. 

\newpage 
We now rewrite the interaction hamiltonian, isolating the contributions of the operators $b_0$ and $b_0^\dagger$ (which may be eventually replaced by $\sqrt{N_0}$). We get
\begin{widetext}
\begin{align}
\begin{split}
\Omega \frac{V}{U}&= 
\frac{1}{2}b_0^\dagger b_0^\dagger b_0b_0
+\frac{1}{2}\sum'_\vk \left( b_0^\dagger b_0^\dagger b_\vk b_{-\vk}+4 b_0^\dagger b_0 b^\dagger_\vk b_\vk + b^\dagger_\vk b^\dagger_{-\vk}b_0b_0\right) +b_0^\dagger b_0 \sum_\vk f^\dagger_\vk f_\vk 
 \\
&~~~+ \sum'_{\vk_1, \vk_2} f^\dagger_{\vk_1} f_{\vk_2} (b^\dagger_{\vk_2-\vk_1}b_0+b_0^\dagger  b_{\vk_1-\vk_2})
+\sum'_{\vk_1\cdots \vk_4} \delta_{\vk_1-\vk_2+\vk_3-\vk_4}\,f^\dagger_{\vk_1} f_{\vk_2}
b^\dagger_{\vk_3} b_{\vk_4} 
  \\
&~~~+\sum'_{\vk_1, \vk_2} \left( b_0^\dagger b^\dagger_{\vk_1+\vk_2} b_{\vk_1}b_{\vk_2}+ b^\dagger_{\vk_1}b^\dagger_{\vk_2} b_{\vk_1+\vk_2}b_0\right)+\frac{1}{2}\sum'_{\vk_1\cdots \vk_4} \delta_{\vk_1-\vk_2+\vk_3-\vk_4}
b^\dagger_{\vk_1} b^\dagger_{\vk_3}b_{\vk_2} b_{\vk_4},
\end{split}
\end{align}
\end{widetext}
where  the prime on the momentum sums indicates that zero momentum boson operators are excluded. 
In the approximations to be considered later, the last two terms will be ignored.  
In the first line, the bosonic terms proportional to  
 $b^{\dagger}_\vk b^{\dagger}_{-\vk}$ or  $b_\vk b_{-\vk}$ are the terms that lead, in the Bogoliubov theory, to the phonon spectrum at small momenta (see Appendix). However, the corresponding modification of the spectrum concerns only a small momentum region, $|\vk| \lesssim k_c$,  with $k_c$  defined by the condition $k_c^2/(2m)=U\rho_b$. 
 As we shall see in the Appendix, the momenta in the relevant loop integrals are of order $k_F$, and the weak coupling condition $U\rho_b\ll \varepsilon_F$ implies $k_c\ll k_F$.
We shall therefore neglect the contribution of the terms $b_\vk b_{-\vk}$ and $b^\dagger_\vk b^\dagger_{-\vk}$ and use the boson spectrum given by the simple mean field approximation, eventually corrected by the boson-fermion interaction.

Finally, we recall that Bose-Fermi mixture can suffer from  instability for some values of the respective densities of bosons and fermions, and a strong enough interaction strength (see e.g.~\cite{Santamore:1, Viverit:1}). Although the present calculations are a priori blind to these instabilities, which occur in channels distinct from the Goldstino channel, we note that the static stability condition of Ref.~\cite{Viverit:1} implies
\begin{align}
\label{eq:stability-condition}
U<\frac{2}{3}\frac{\varepsilon_F}{\rhof}\equiv U_{c1}
\end{align}
 in the current supersymmetric setup.
In terms of the scattering length, this condition amounts to $\kf a<\pi/2\simeq 1.6$.\\

\section{Explicit calculation at weak coupling}
\label{sec:spectrum}

We turn now to the explicit evaluation of the Goldstino propagator in the weak coupling case. 

\subsection{Free case ($U=0$)}
\label{ssc:U=0}

\begin{figure}[h] 
\begin{center}
\includegraphics[width=0.3\textwidth]{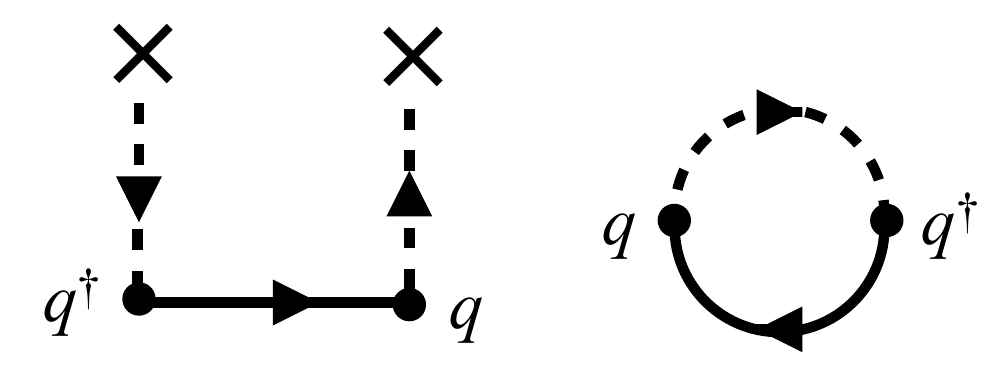} 
\caption{The one-loop diagrams contributing to ${G}^{0}$.
The full (dashed) line represents a fermion (boson) propagator. 
The convention used for these diagrams, and those below that contain arrows is as follows. 
The time flows from left to right. 
An arrow pointing to the right indicates a ``particle'', while an arrow pointing to the left indicates a ``hole''. 
The boson hole propagator is disconnected, and represented by the diagram on the left, where  the dashed lines terminated by a cross denote an expectation of $b_0$ if the arrow points away from the cross and $b_0^\dagger$ if the arrow points towards the cross.
It corresponds to the first term of the first line of Eq.~(\ref{eq:Gtil-U=0}) while the diagram on the right corresponds to the second term.  } 
\label{fig:oneloop} 
\end{center} 
\end{figure}

We start the analysis with the non interacting limit, $U=0$.
In this case, $G$ is given by the one-loop diagrams in Fig.~\ref{fig:oneloop}, whose evaluation yields
\beq
\label{eq:Gtil-U=0}
G^0(p) 
&=& -\frac{\rho_b(1-n_\vp)}{\bar\omega -\epsilon^0_\vp}
-\frac{1}{\Omega}\sum_\vk \frac{n_\vk+n_\vp N_0 \delta_{\vp+\vk}}
{\bar\omega+\epsilon^0_{\vk+\vp}-\epsilon^0_\vk},\nn
&=& -\frac{\rho_b}{\bar\omega -\epsilon^0_\vp}
-\frac{1}{\Omega}\sum_\vk \frac{n_\vk}
{\bar\omega+\epsilon^0_{\vk+\vp}-\epsilon^0_\vk},\
\eeq
where $\epsilon^0_{\vk}\equiv \vk^2/(2m)$,   $\bar\omega=\omega+\Delta\mu_0$ with $\Delta\mu_0=k_F^2/(2m)$,  and $N_0=\Omega\rho_b$, while $n_\vp\equiv \theta(\kf-p)$ denotes the fermion occupation number.

\begin{figure}[t] 
\begin{center}
\includegraphics[width=0.4\textwidth, angle=0]{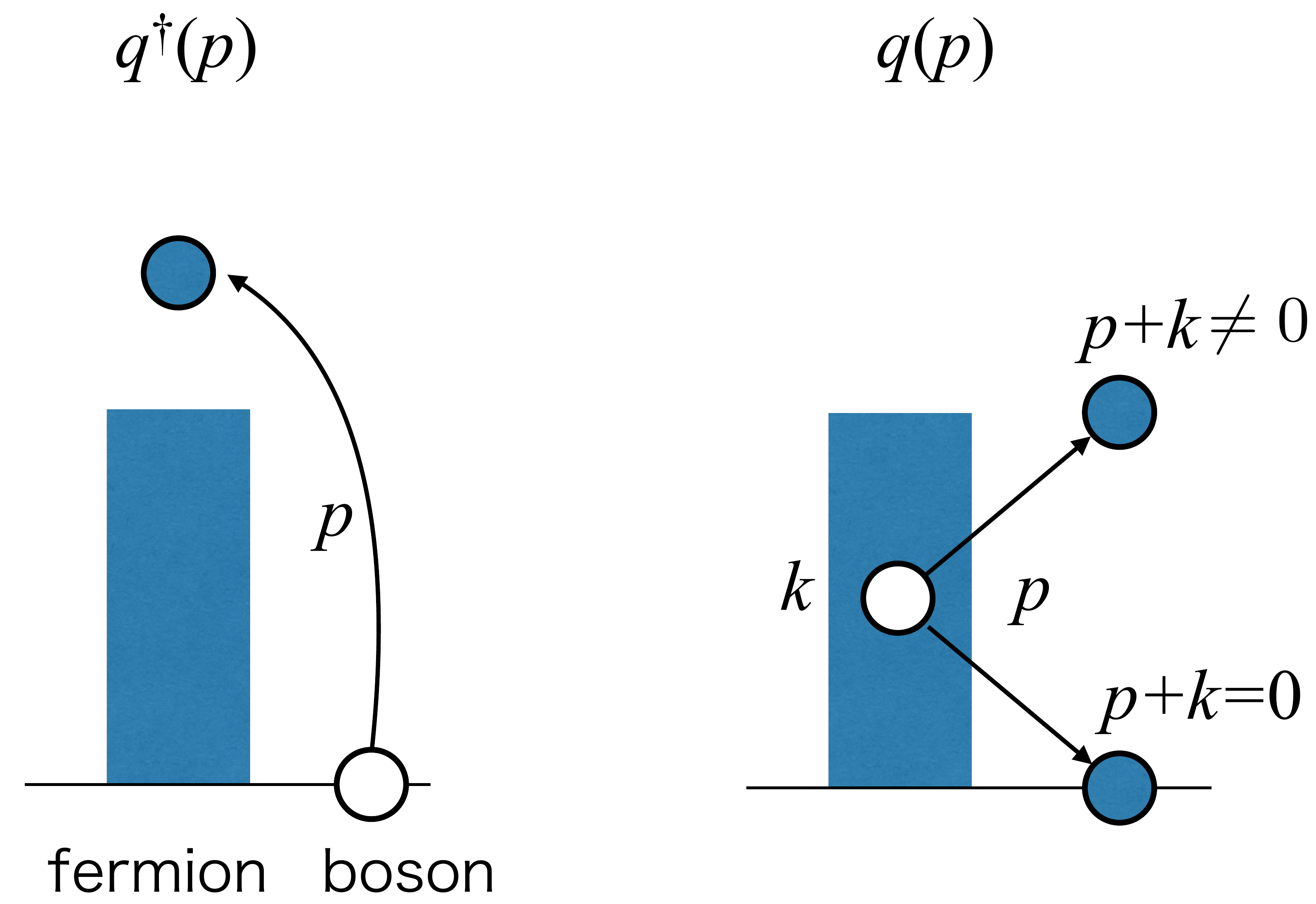} 
\caption{Particle-hole excitations contributing to the Goldstino.
The process in the left represents $q^\dagger$, which replace a boson with a fermion, while the one in the right represents $q$, which replaces a fermion with a boson.
The blue square represents the Fermi sea.
\label{fig:particle_hole} 
} 
\end{center} 
\end{figure}

\begin{figure}[h] 
\begin{center}
\includegraphics[width=0.5\textwidth]{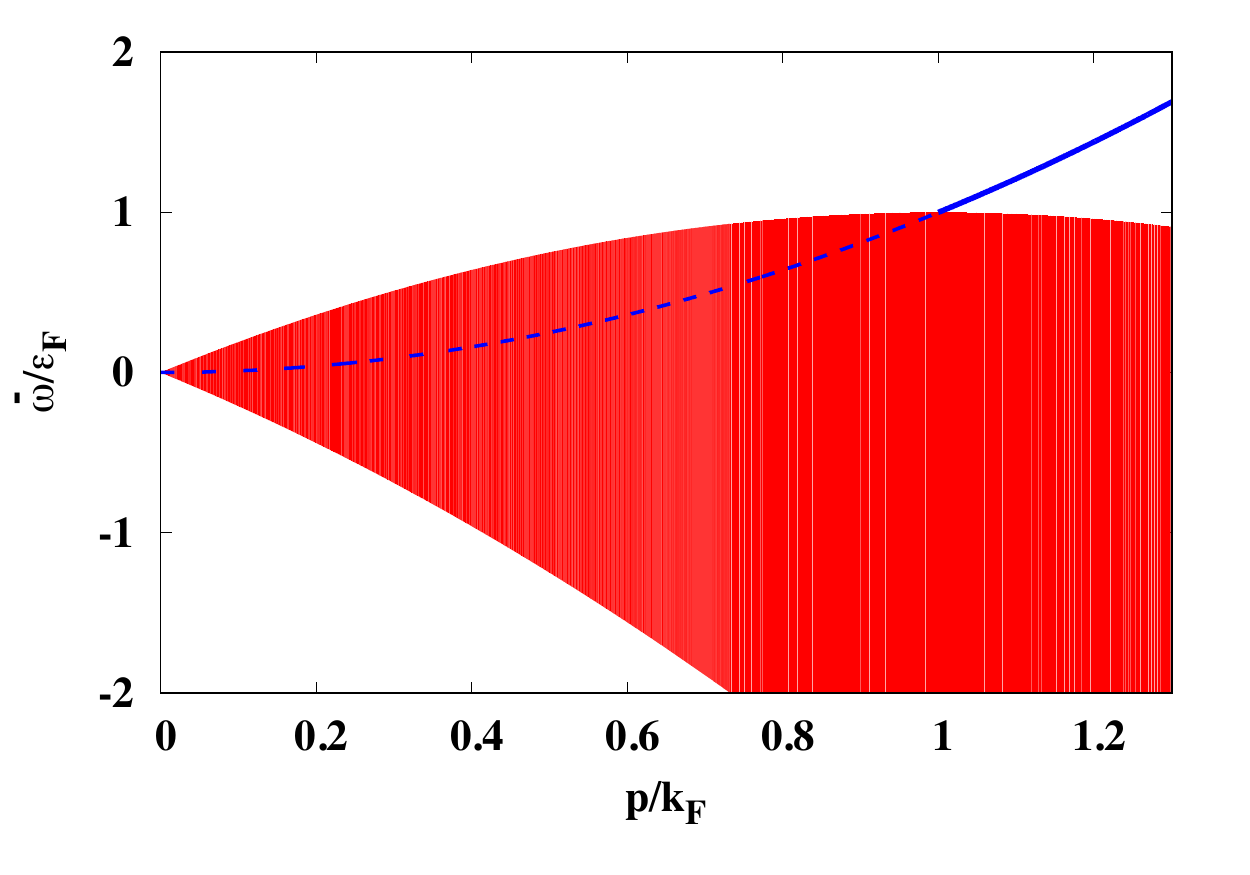} 
\caption{The continuum (red shaded area) and the pole (blue dashed and solid line) in $\sigma_{{G}}$.
The pole leaves the continuum at $|\vp|=k_F$.
} 
\label{fig:range-U=0} 
\end{center} 
\end{figure}

\begin{figure}[t] 
\begin{center}
\includegraphics[width=0.5\textwidth]{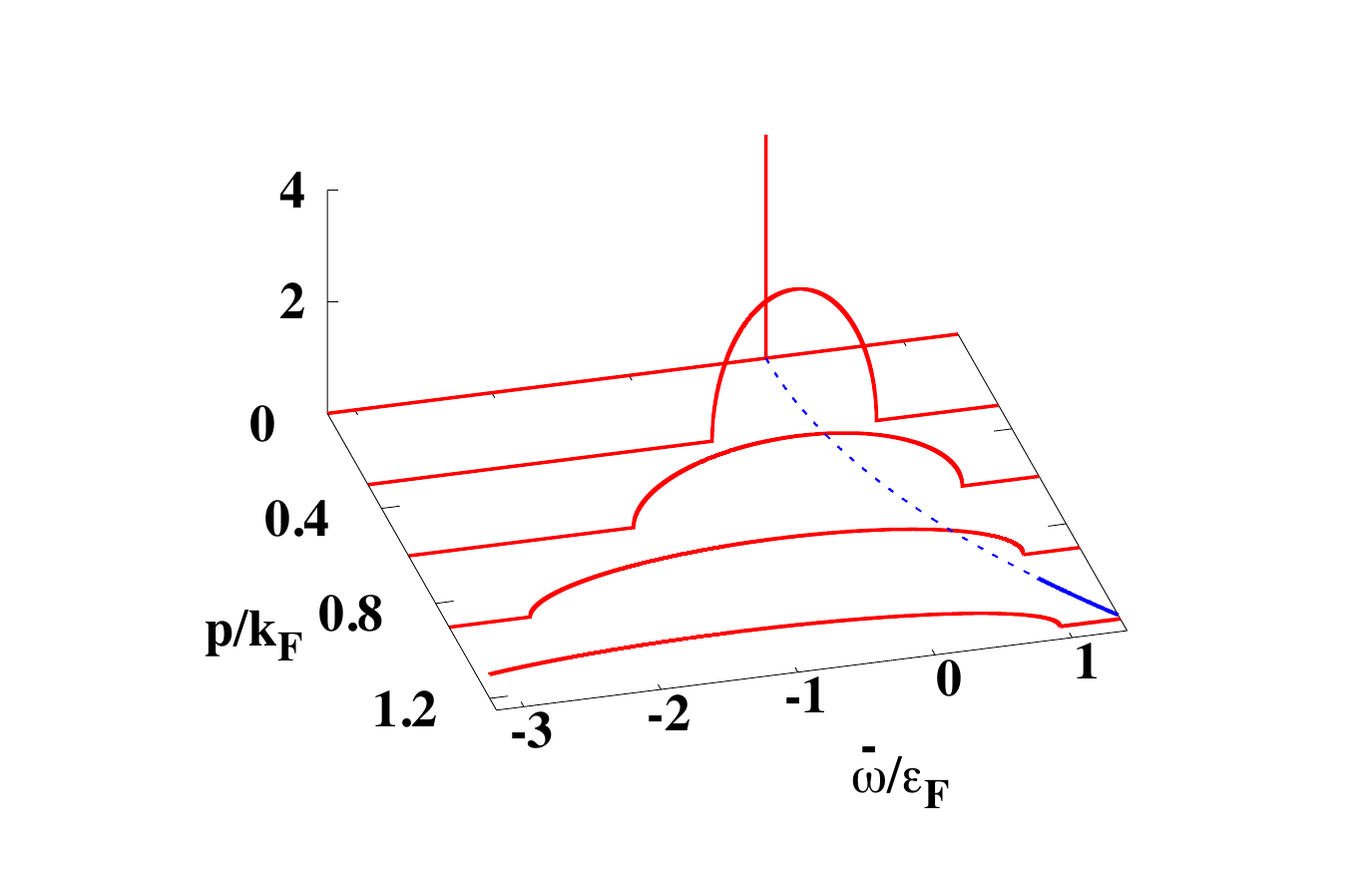} 
\caption{Color online. The spectral function of the Goldstino in the non-interacting case. 
The full (red) line represents the continuum contribution, Eq.~(\ref{eq:spectle-Gtil-U=0}), which is peaked at $\bar\omega=-\vp^2/(2m)$. 
The dotted (blue) line represents the location of the pole, $\bar\omega=\vp^2/2m$,  Eq.~(\ref{eq:spectle-Gtil-U=0pole}). 
The pole out of the continuum is plotted with the full (blue) line.
As the momentum decreases, the peak in the continuum sharpens and eventually turns into a delta function $\delta(\bar\omega)$ at $\vp=\vzero$.
The unit of $\sigma_G$ is $\rhof/\eF$.
} 
\label{fig:spectal0} 
\end{center} 
\end{figure}

The second contribution in the first line of
Eq.~(\ref{eq:Gtil-U=0}) corresponds to excitations induced by $q(\vp)$, which produces a hole with momentum $\vk$ in the Fermi sea, turning the corresponding fermion into a boson with momentum $\vk+\vp$ (see Figs.~\ref{fig:oneloop} and \ref{fig:particle_hole},  right). The corresponding poles lie at $-\omega=E_m-E_0+\Delta\mu_0=\epsilon^0_{\vk+\vp}-\epsilon^0_{\vk}+\Delta\mu_0=
 \vp^2/(2m)+ \vp\cdot\vk/m+k_F^2/(2m) \ge 0 $. 
These excitations form actually a continuum in the range  
\begin{align}
\label{eq:continuum-range-U=0}
-\frac{\vp^2}{2m}-\frac{k_F |\vp|}{m}<\bar{\omega}<-\frac{\vp^2}{2m}+\frac{k_F |\vp|}{m}.
\end{align} 
The corresponding contribution to the Goldstino spectral function reads
\begin{align}
\label{eq:spectle-Gtil-U=0}
\begin{split}
\sigma_{\rm cont}(p) 
&= \frac{m}{4\pi|\vp|}(\kf^2-\kcf^2)\theta(\kf-\kcf), 
\end{split}
\end{align}
where $\kcf\equiv m|\bar{\omega}+\vp^2/(2m)|/|\vp|$. The existence of the continuum is directly related to the presence of a Fermi sea, and its support is indeed directly related to the magnitude of the Fermi momentum $k_F$.  
The range of this continuum is displayed in  Fig.~\ref{fig:range-U=0}, and its shape is in Fig.~\ref{fig:spectal0}. 
We have used there  $\kf$ ($\varepsilon_F$) as a unit of momentum (energy).

The first term in the first line of Eq.~(\ref{eq:Gtil-U=0}) corresponds to excitations induced by $q^\dagger_\vp$, which turns a  boson in the condensate (with momentum $\vk=\vzero$) into a fermion above the Fermi sea, i.e. with momentum $|\vp| \ge k_F$ (see Figs.~\ref{fig:oneloop} and \ref{fig:particle_hole},  left). 
There exists  another pole contribution hidden in the second term of  the first line of Eq.~(\ref{eq:Gtil-U=0}): it corresponds to a hole in the Fermi sea with a momentum $\vp$ such that the associated boson fills the condensate (see Fig.~\ref{fig:particle_hole}, right). Such a process is amplified by the presence of the condensate, hence the factor $N_0$ accompanying this excitation.   
This particular contribution is cancelled by  the term proportional to $n_\vp$ in the first term in the first line of Eq.~(\ref{eq:Gtil-U=0}). 
The net result is  the first term of the second line of Eq.~(\ref{eq:Gtil-U=0}), which yields the 
following contribution to the  spectral function 
 \beq\label{eq:spectle-Gtil-U=0pole}
 \sigma_{\rm pole}(p)=2\pi \rho_b
 \delta\left(\bar{\omega}-\frac{\vp^2}{2m}\right).
 \eeq
The pole position is displayed in Fig.~\ref{fig:range-U=0}.
  Aside from the factor $\rho_b$, which reflects the degeneracy of the condensate,  this spectral function is that of a free fermion, with the associated dispersion relation, $\omega=\vp^2/(2m)-\Delta\mu_0$. 
This is natural since the corresponding excitations involve adding or removing a particle  in the condensate, which costs no energy/momentum. 
  
 The behavior of the Goldstino spectral function is illustrated  in Fig.~\ref{fig:spectal0}. More precisely, what is plotted in Fig.~\ref{fig:spectal0} is the continuum contribution, corresponding to the second line of Eq.~(\ref{eq:Gtil-U=0}), that is $\sigma_{\rm cont}(p) $ given by Eq.~(\ref{eq:spectle-Gtil-U=0}). This continuum  contributes as a pole at $ |\vp| =0$, whose location at $\bar{\omega}=0$ coincides with that  in Eq.~(\ref{eq:spectle-Gtil-U=0pole}), and whose residue is $\rho_f$.  At finite momentum,  this turns into a peak that broadens as $ |\vp|$ increases.   
 In addition to the peak in the continuum, the location of the pole, with constant residue $\rhob$, is also indicated (see Eq.~(\ref{eq:spectle-Gtil-U=0pole})).  
 When $ |\vp|> k_F$, this pole  is out of the region occupied by the continuum. 
 Note that 
at this level, the spectral function is  essentially the same as in the case without BEC~\cite{Blaizot:2015wba}. 

At $|\vp|=0$, Eq.~(\ref{eq:Gtil-U=0}) reduces to ${G}^0(\omega,\vzero) = - \rho/{\bar{\omega}}$, which exhibits a pole at $\bar\omega=0$ with residue equal to the total density. This is in agreement with the general expression (\ref{eq:G0-p=0-pole}), which is a consequence of the underlying supersymmetry.  Note that at $\vp=\vzero $ only one of the two types of processes displayed in Fig.~\ref{fig:particle_hole} contribute. To see that, it is useful to consider how things evolve as we change $\rho_f$ keeping $\rho$ fixed. When $\rho_f=0$, only boson hole excitations are allowed, and their degeneracy 
 is proportional to $\rho_b=\rho$. 
 As soon as $\rho_f\ne 0$ however, these excitations are blocked and replaced by fermion hole excitation with zero momentum. 
 This transition is amplified by the Bose enhancement factor and hence its contribution is proportional to $\rho_b$. 
In addition, there are the excitations involving a fermion hole and a boson particle with nonzero and identical momenta, which contribute to the residue a factor proportional to $\rho_f$. 
 In summary, as soon as $\rho_f\ne 0$, only the process displayed in the right part of Fig.~\ref{fig:particle_hole} contributes with the factor $\rhof+\rhob=\rho$.

The nature of the Goldstino is particularly simple in the two  limits where $\rho_f=0$ or $\rho_b=0$. In the first case, $\rho_f=0$ and $\Delta\mu_0=0$. The system is therefore supersymmetric. The Goldstino propagator is given by the first term of the second line of Eq.~(\ref{eq:Gtil-U=0}).
The Goldstino in that case is like a free fermion excitation. 
In the other limit, $\rho_b=0$, supersymmetry is explicitly broken by the non vanishing value of $\Delta\mu_0$. In addition, the Goldstino exists as a pole  only at $\vp=\vzero$. 
The Goldstino pole at $\bar\omega=0$ for $\vp= \vzero$, with strength $\rho_f$,  turns into  a branch cut singularity,  corresponding to the continuum of finite momentum fermion hole excitations, as illustrated in Figs.~\ref{fig:range-U=0}  and \ref{fig:spectal0}.
However, we will see that this behavior is completely changed if one considers the effect of interactions. \\


\subsection{Interacting case ($U\neq0$)}
\label{ssc:U-finite}

Let us now proceed to the interacting case, $U\ne 0$, and focus on the leading order at weak coupling. We first analyze how the chemical potentials and  the single particle energies are modified by mean field effects.  The mean field approximation corresponds to the following effective interaction hamiltonian
\begin{align}\label{meanfieldham}
\begin{split}
\frac{V}{U}&= 
\rho_b\rho_f \Omega+ \rho_b^2\frac{\vol}{2}+\rho_b\sum_\vk :f^\dagger_\vk f_\vk: \\
&~~~ +\rho_f \sum^\prime_\vk
b^\dagger_{\vk} b_{\vk} 
+\frac{\rho_b}{2}\sum'_\vk 4 b^\dagger_\vk b_\vk  ,
\end{split}
\end{align}
where the normal ordering of the fermion operator is with respect to the non interacting Fermi sea ($\langle \sum_\vk :f^\dagger_\vk f_\vk:\rangle=0$). 
The first two terms in this expression are the expectation value of the interaction terms for the noninteracting ground state.
After adding the kinetic energy of the fermions, this expression of the ground state energy can be used to determine the chemical potentials, from the usual relation,  $\mu_i=\partial \langle H\rangle/\partial N_i$. In the present mean field approximation,   $\langle H\rangle=\langle H_f\rangle+U[\Nf N_b+N_b^2/2]/\Omega$, where $\langle H_f\rangle/\Omega=3\rhof\varepsilon_F/5$ denotes the kinetic energy of the Fermi sea. By taking the derivatives with respect to $N_f$ and $N_b$, one gets
\beq
\label{eq:mus-MF-T=0}
\muf= \varepsilon_F+U\rhob,\qquad 
\mub= U\rho,
\eeq
so that $\Delta\mu= \varepsilon_F-U\rhof$. Note that $\Delta \mu$, responsible for the explicit breaking of supersymmetry, vanishes when the fermion density vanishes or $U=U_{c2} \equiv \varepsilon_F/\rhof$.
This value of $U$ is translated to $\kf a=3\pi/4\simeq 2.4$.

\begin{figure}[t] 
\begin{center}
\includegraphics[width=0.45\textwidth]{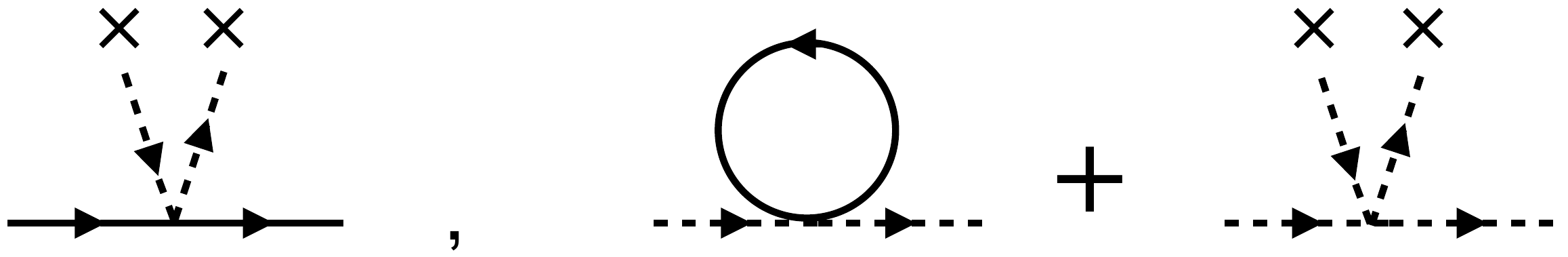} 
\caption{The diagrams contributing to the mean field correction to the fermion (full line)  and the boson (dashed line) excitation energies.
The cross attached to a dashed line represents the contribution of the condensate $\sqrt{\rho_b}$. 
We ignore the anomalous contributions to the boson self-energy. 
Note that the last diagram contributes a correction $2\rhob$ to the energy of a boson outside the condensate and only $\rhob$ to the energy of a boson in the condensate.
} 
\label{fig:MF-spectrum} 
\end{center} 
\end{figure}

The corrections to the single particle energies can be read off the effective hamiltonian (\ref{meanfieldham}). That of the fermion is  coming from the left diagram in Fig.~\ref{fig:MF-spectrum}, while that of the boson energy comes from the last two diagrams.
The contribution of these diagrams  amounts to a simple shift of the single particle energies of the fermion ($\delta\epsilon^f$) and the boson ($\delta\epsilon^b$) given by  
\beq
\label{eq:MF-dispersion}
\delta\epsilon^f=  U\rhob,\qquad 
\delta\epsilon^b= U(\rhof+2\rhob). 
\eeq
 Note that, in the case of bosons, the correction  above applies only to particles with non zero momentum. For a particle in the condensate, the correction is 
\beq\label{eq:MF-dispersion2}
\delta\epsilon^b_0= U(\rhof+\rhob).
\eeq
The missing factor 2 in Eq.~(\ref{eq:MF-dispersion2}), as compared to Eq.~(\ref{eq:MF-dispersion}), reflects the absence of an exchange term in the former case. As a result, particles in the condensate experience less repulsion than particles outside the condensate. This generates a gap in the boson spectrum, which would  disappear if  phonons were taken into account (see Appendix). \\

In the mean field (MF) approximation, $G$ is still given by the one-loop diagrams in Fig.~\ref{fig:oneloop}, with the MF single particle energies. It is again convenient to separate the contribution from the pole from that of the continuum. The contribution from the continuum reads
\beq
\label{eq:Gtil-MF0cont}
\GMFcont (p) 
&=& -\frac{1}{\Omega}\sum'_\vk \frac{n_\vk}
{\bar\omega+\epsilon^b_{\vk+\vp}-\epsilon^f_\vk}\nn &=&-\frac{1}{\Omega}\sum'_\vk \frac{n_\vk}
{\bar\omega+U\rho+\epsilon^0_{\vk+\vp}-\epsilon^0_\vk}, 
\eeq
and that of the pole is
\beq
\label{eq:Gtil-MF0pole}
\GMFpole (p) =
-\frac{\rho_b}{\bar\omega -\epsilon^0_\vp+U\rho_f},
\eeq
where we have used Eqs.~(\ref{eq:MF-dispersion}) and (\ref{eq:MF-dispersion2}) for the single particle energies, and here $n_\vk=\theta(\mu_f-\epsilon^f_\vk)$. 
The shifts in the particle-hole energies are different in Eqs.~(\ref{eq:Gtil-MF0cont}) and (\ref{eq:Gtil-MF0pole}) ($U\rho$ versus $U\rhof$). This is  because the boson appearing in the former equation is not in the condensate while it is in the latter equation, and the corresponding boson self-energies differ.

As can be seen from the previous formulae, the  pole structure is affected by the mean field corrections to the single particle energies (except when $\rho_f=0$). In the case of $\rho_b=0$, for instance,  the Goldstino pole is shifted away from $\bar\omega=0$ by an amount $U\rho_f$. 
However, the mean field approximation is not consistent in this case, and 
particle-hole interactions need to be taken into account. These are generated by  the following two terms in the effective hamiltonian
\beq\label{phinteraction}
&&H_3= \frac{U}{\Omega}\sum'_{\vk_1, \vk_2} f^\dagger_{\vk_1} f_{\vk_2} (b^\dagger_{\vk_2-\vk_1}b_0+ b_0^\dagger b_{\vk_1-\vk_2}) , \nn
&&H_{4}=\frac{U}{\Omega}\sum'_{\vk_1\cdots \vk_4} \delta_{\vk_1-\vk_2+\vk_3-\vk_4}\,f^\dagger_{\vk_1} f_{\vk_2}
b^\dagger_{\vk_3} b_{\vk_4} .  \nn
\eeq
In order to study the effects of these interactions, we shall examine first the cases where one of the two densities vanishes. In these cases, the two contributions to the Goldstino propagator decouple, which makes the analysis simpler. Then we consider the general case. 

\subsubsection{Case $\rho_b=0$}

\begin{figure}[t] 
\begin{center}
\includegraphics[width=0.5\textwidth]{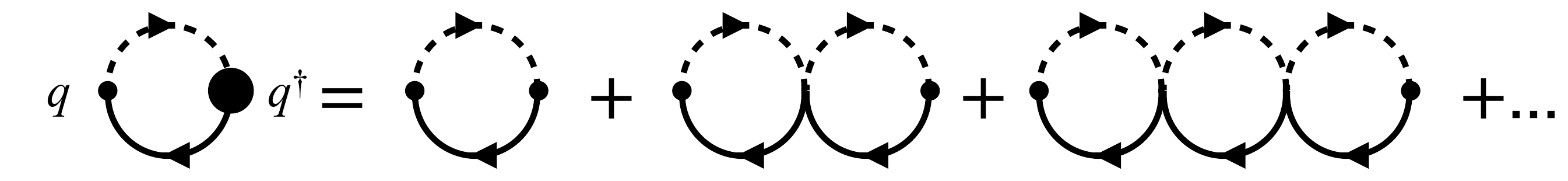} 
\caption{The ring diagrams that are summed in the RPA calculation of $\GRPA$, Eq.~(\ref{eq:Gtil-RPA}). Note that the propagators are full mean field propagators. The interaction joining two successive bubbles is the one in the second line of Eq.~(\ref{phinteraction}), namely $H_4$.}
\label{fig:RPA} 
\end{center} 
\end{figure}

When $\rho_b=0$, the one-loop contribution reduces to the continuum contribution, Eq.~(\ref{eq:Gtil-MF0cont}) with $U\rho\to U\rho_f$. At $|\vp|=0$, the pole lies  at $\bar\omega=-U\rho_f$, corresponding to the shift in the particle-hole continuum, $U\rho_f=\delta\epsilon_b-\delta\epsilon_f$.
However, as we have already indicated, the mean field (one-loop) approximation is in this case not consistent. Indeed, when we analyze the small $\bar{\omega}$ region, one finds that additional diagrams  contribute with the same order of magnitude as the one-loop diagram :
From the expression (\ref{eq:Gtil-MF0cont}) above, we see that $\GMFcont (\omega,\vp=\vzero)\sim U^{-1} $ when $\bar{\omega}\ll U\rho_f$.
The diagram with two rings drawn in Fig.~\ref{fig:RPA} is of order $U\times  U^{-2} \sim U^{-1}$, where $U$ comes from the vertex and $U^{-2}$ from the two rings.
Thus this two-ring diagram has the same order of magnitude as the one-loop diagram.
The same result is obtained also for the ring diagrams containing more loops, so all of them need to be summed in order to get a correct result~\cite{Blaizot:2015wba, Shi:2009ak}.
Such a resummation, commonly referred to as  the random phase approximation (RPA),  yields the following expression for the Goldstino propagator
\beq
\label{eq:Gtil-RPA}
\GRPA (p) 
= \frac{1}{  [\GMFcont (p)]^{-1}+U  },
\eeq
which, at zero momentum, reduces to
\beq
\label{eq:Gtil-RPA-p=0}
\GRPA (\omega,\vzero) 
= -\frac{\rhof}{\bar{\omega}}.
\eeq
This is now the correct result (\ref{eq:G0-p=0-pole}). This provides a beautiful illustration of the collective nature of the Goldstino in this case: the degenerate particle-hole excitations (a fermion hole and a boson particle with the same momenta, producing an excitation with energy  $\bar\omega=-U\rho_f$) being pushed up at $\bar\omega=0$ by the particle-hole interaction.  
In other words, the particle-hole interaction cancels the shift of the single particle energies, thereby shifting back the excitation to $\bar\omega=0$, as expected from symmetry considerations.  Furthermore, the shift caused by the particle-hole interaction pushes the Goldstino out of the continuum at finite momentum, hindering its natural broadening as the momentum increases.  

At small $|\vp|$,  the Goldstino then appears as a single pole:
To find the corresponding dispersion relation  $\bar\omega(\vp)$,   we  expand Eq.~(\ref{eq:Gtil-MF0cont}) in the vicinity of the Goldstino pole, i.e., in powers  of $|\vp|$ and $\bar{\omega}$. We obtain, up to quadratic order, 
\begin{align}\label{Gexpanded}
\begin{split}
\GMFcont(p)
&\simeq -\frac{1}{U}
\left(1+a\bar{\omega}+b\frac{\vp^2}{2m} +a^2\bar{\omega}^2+c\bar{\omega}\frac{\vp^2}{2m}
\right),\\
\end{split}
\end{align}
where the coefficients $a,b,c$ are given by 
\begin{align}
\begin{split}
a&= -\frac{1}{U\rhof},~~
b= \frac{1}{U\rhof}\left(-1+\frac{4}{5}\frac{\varepsilon_F}{U\rhof}\right),\\
c&= \frac{2}{(U\rhof)^2}\left(1-\frac{6}{5}\frac{\varepsilon_F}{U\rhof}\right).
\end{split}
\end{align}
The resulting expression for the Goldstino propagator reads
\begin{align}
\begin{split}
\GRPA (p) 
&\simeq -\frac{Z_{{G}}}{\bar{\omega}-\alpha_f \vp^2/2m} , 
\end{split}
\end{align}
where 
\begin{align}
\label{eq:Z-Gtilde}
Z_{{G}}&\equiv \rhof \left[1
-\frac{4}{5}\left(\frac{|\vp|}{\kf} \frac{\eF}{U\rhof} \right)^2   \right],\\
\label{eq:alphaf}
\alpha_f&\equiv -1+\frac{4}{5}\frac{\varepsilon_F}{U\rhof} .
\end{align}
We note that the dispersion relation is quadratic, but the coefficient $\alpha_f$ differs from the coefficient $\alpha_s$ in Eq.~(\ref{eq:sumrule2-G}). In contrast to $\alpha_s$, which in the present case contributes to the term $-1$ in Eq.~(\ref{eq:alphaf}),  $\alpha_f$ depends on the interaction strength and it can turn positive for small enough values of $U$.

\subsubsection{Case $\rho_f=0$}

The situation is different in the case  $\rho_f=0$, to which we now turn. Then, only the pole term contributes, that is
\begin{align}
\label{eq:Gtil-MF2}
\begin{split}
G^{\mbox{\tiny MF}}_{\rm pole}(p) 
&= -\frac{\rho_b}{\bar\omega -\epsilon^0_\vp}.
\end{split}
\end{align}
This is identical to the expression obtained without interactions: the energy of a fermion particle receives a mean field correction, which is identical to that of the energy of a particle of the condensate, and both corrections cancel in the denominator.  
In this case the one-loop approximation gives the correct result.  

There is therefore a profound asymmetry between the two types of excitations. In the first case, when $\rho_b=0$, the Goldstino appears as a superposition of particle-hole excitations (with the hole being a fermion), and the particle-hole interaction plays an essential role in the emergence of the collective mode. In the second case, $\rho_f=0$, the hole is a boson in the condensate, and the Goldstino appears as a fermion propagating in the field of the condensate: its properties are entirely captured by the mean field approximation. \\

When both densities are finite, the two types of excitations couple. We shall now analyze the effect of this coupling. For simplicity, we shall focus first on the case $\vp=\vzero$.

\subsubsection{General case, $\vp= \vzero$}

\begin{figure}[t] 
\begin{center}
\includegraphics[width=0.35\textwidth,angle=0]{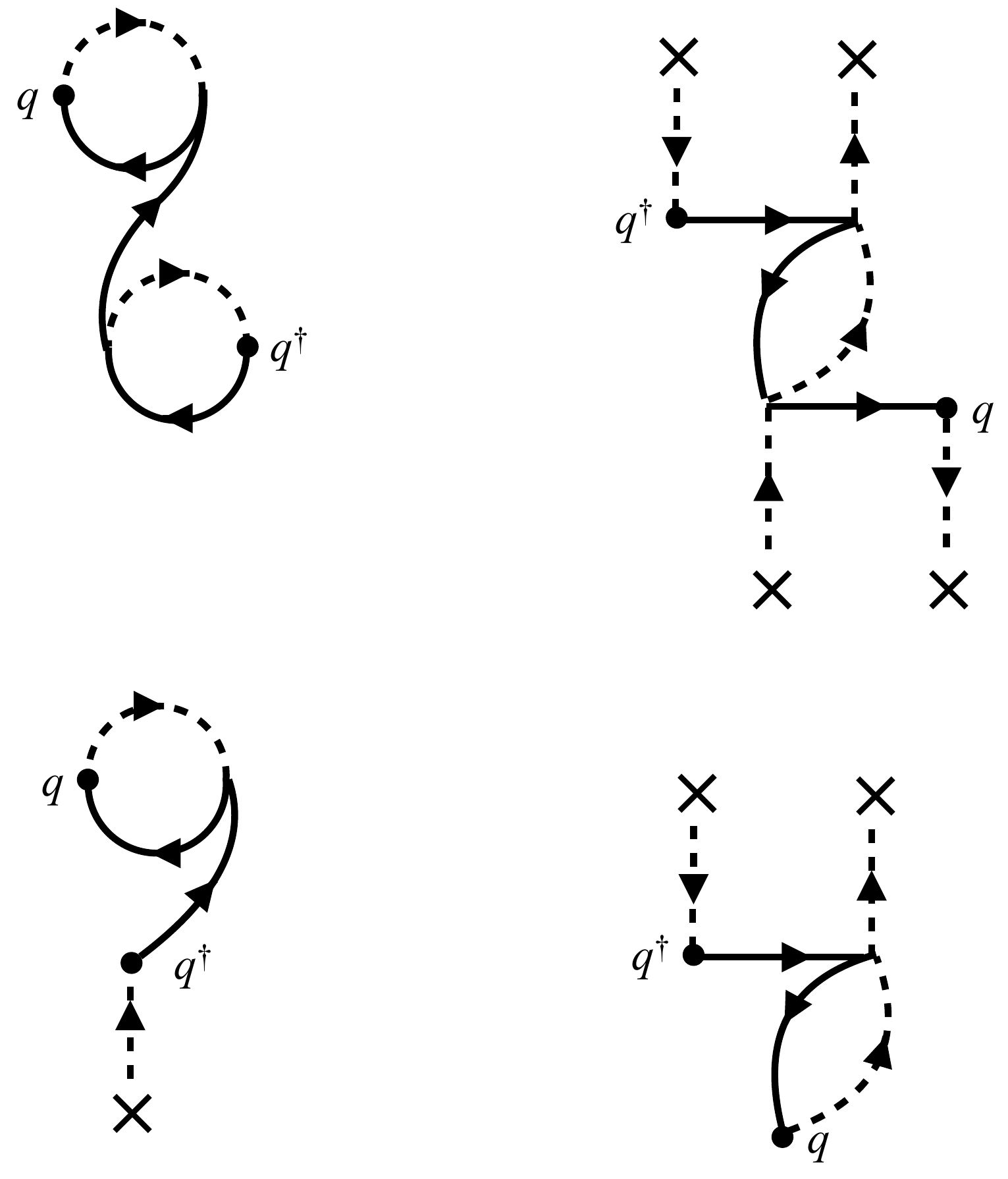} 
\caption{The new types of diagrams that are allowed by the interaction. In these diagrams, the bubble is the mean field bubble $\GMFcont (p) $, Eq.~(\ref{eq:Gtil-MF0cont}), while the single line is the propagator $\GMFpole (p)$, Eq.~(\ref{eq:Gtil-MF0pole}). The single bubble can be replaced by the RPA bubble sums, while the diagrams on the top can be iterated. 
} 
\label{fig:newdiagrams} 
\end{center} 
\end{figure}

In the general case, the two types of processes mix, thanks to the fermion-boson interaction that involves one particle in the condensate, namely the term $H_3$ in Eq.~(\ref{phinteraction}).   The mixing arises from the fact that the interaction can couple a fermion hole and a fermion particle. New types of diagrams such as the ones in Fig.~\ref{fig:newdiagrams} then appear.
That there is a need for extra contributions can be seen from the following argument. Consider again the RPA. 
For general $\rhof$ and $\rhob$, Eq.~(\ref{eq:Gtil-RPA-p=0}) becomes
\beq
\label{eq:Gtil-RPA-p=0b}
\GRPA (\omega,\vzero) 
= -\frac{\rhof}{\bar{\omega}+U\rho_b}.
\eeq
Thus when $\rho_b\ne 0$, the pole of $G^{\mbox{\tiny RPA}}$ is no longer at $\bar \omega=0$, which is in conflict with the symmetry argument. 
We shall see that the problem is cured by extra contributions to which we now turn. 

For a systematic classification of the diagrams, we decompose the Goldstino Green's function into three components, $\tilde{G}$, $G_3$, and $G_S$. 
Such a decomposition  naturally emerges if in the expression of the supercharge operators $q(\vp)$ and $q^\dagger(\vp)$ (see Eq.~(\ref{superchargep})), we replace the zero momentum boson  operators $b_0^\dagger$ and $b_0$ by shifted operators according to Eq.~(\ref{eq:BEC-boson-decompose}). The three components  $\tilde{G}$, $G_3$, and $G_S$ correspond then to contributions to the Goldstino propagator that are respectively independent, linear or quadratic in $b_0^\dagger$ and $b_0$.
Furthermore, from now on, we shall focus on the topology of the new diagrams without distinguishing the difference between particle and  hole by arrows.

\begin{figure}[t] 
\begin{center}
\includegraphics[width=0.45\textwidth,angle=0]{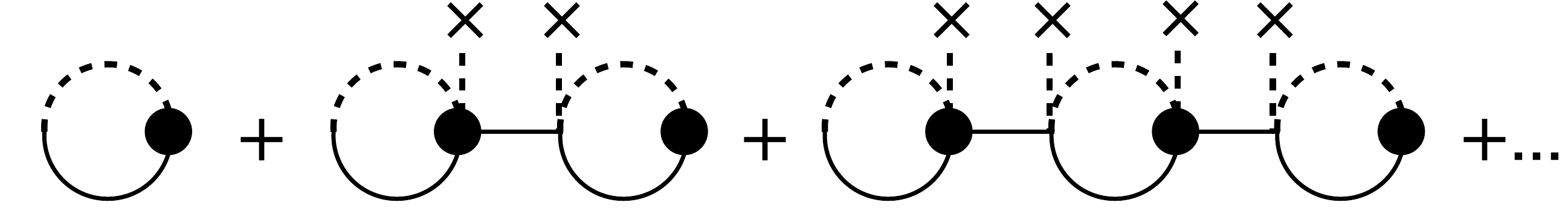} 
\caption{The diagrams containing the mixing between between the RPA diagrams ($\GRPA$) and $\GMFpole$ contributing to $\tilde{G}$.
The blob represents the RPA diagrams.
} 
\label{fig:1PR} 
\end{center} 
\end{figure}

The first contribution, $\tilde{G}$, corresponds to diagrams whose two ends are connected with bosons  with finite momenta (i.e. a boson particle, not a particle of the condensate). 
In the special case, $\rhob=0$, this coincides  with the full  Goldstino propagator $\GRPA$.
In the more general case, it is corrected by the  diagrams in Fig.~\ref{fig:1PR}. 
The second diagram in this figure is of order $U^2\times U^{-2}\times U^{-1}$, where the factor $U^2$ comes from the two vertices, the factor $U^{-2}$ from the two bubbles, and $U^{-1}$ from the mean field propagator near $\bar\omega=|\vp|=0$, i.e., 
\beq
\label{eq:Gtil-MF0pole2}
\GMFpole (p) = -\frac{\rho_b}{\bar\omega-\epsilon_\vp^0+U\rho_f} \approx -\frac{\rho_b}{U\rho_f}. 
\eeq
This diagram has the same order of magnitude $\sim U^{-1}$ as the RPA diagrams, and the same holds for the entire family of diagrams  displayed in Fig.~\ref{fig:1PR}.  
Their sum yields 
 \begin{align}
\tilde{G}(p) 
&= \frac{1}{[\GRPA (p)]^{-1}-U^2\GMFpole (p)},
\end{align}
where $\GRPA (p)$ is given by Eq.~(\ref{eq:Gtil-RPA}).
At zero momentum, it reduces to 
 \begin{align}
 \label{eq:Gtil-1PR-p=0}
\tilde{G}(\omega,\vzero) 
&= -\left[\frac{\rhof^2}{\rho}\frac{1}{\bar{\omega}}
+\frac{\rhof\rhob}{\rho}\frac{1}{\bar{\omega}+U\rho}\right].
\end{align}
Here we have one pole with no gap, and another one with a finite gap ($\bar{\omega}=-U\rho$), whose existence is due to the presence of a BEC.
One may interpret this result in terms of level repulsion:
Before the mixing caused by the  diagrams of Fig.~\ref{fig:newdiagrams}, the poles were located at $\bar{\omega}=-U\rhob$  for $\GRPA$ (Eq.~(\ref{eq:Gtil-RPA-p=0b})), and at  $\bar{\omega}=-U\rhof$  for  $\GMFpole $ (Eq.~(\ref{eq:Gtil-MF0pole2})).
The mixing  causes a repulsion between these two poles, leading eventually to the result just mentioned. Of course the supersymmetry plays an important role here in insuring for instance that the mixing yields a Goldstino pole at $\bar\omega=0$. 

\begin{figure}[t] 
\begin{center}
\includegraphics[width=0.48\textwidth]{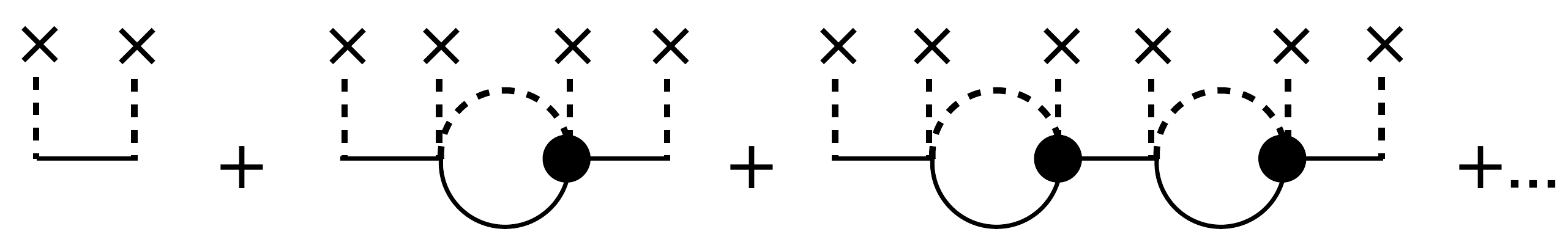} 
\caption{The diagrams containing the mixing between the RPA diagrams ($\GRPA$) and $\GMFpole$ contributing to $G_S$. 
} 
\label{fig:1PR-S} 
\end{center} 
\end{figure}

A similar phenomenon is observed for the other components of the Goldstino propagator. 
We consider now $G_S$,  the component whose two ends are connected with the condensate. 
In the special case, $\rhof=0$, it agrees with the total Goldstino propagator given in Eq.~(\ref{eq:Gtil-MF2}). 
The corrections at general densities are given by  the  diagrams shown in Fig.~\ref{fig:1PR-S}. Following the same reasoning as for $\tilde G$, one can see that these diagrams  contribute with the same order of magnitude when $\bar{\omega}$ and $|\vp|$ are small.
The resulting expression reads
  \begin{align}
   \label{eq:S-1PR}
G_S(p) 
&= \frac{1}{[\GMFpole (p)]^{-1} -U^2 \GRPA(p)}.
\end{align}
At zero momentum, this reduces to an expression very similar to Eq.~(\ref{eq:Gtil-1PR-p=0}), viz.
  \begin{align}
   \label{eq:S-1PR-p=0}
G_S(\omega,\vzero) 
&=  -\left[\frac{\rhob^2}{\rho}\frac{1}{\bar{\omega}}
+\frac{\rhof\rhob}{\rho}\frac{1}{\bar{\omega}+U\rho} \right].
\end{align}

\begin{figure}[t] 
\begin{center}
\includegraphics[width=0.48\textwidth]{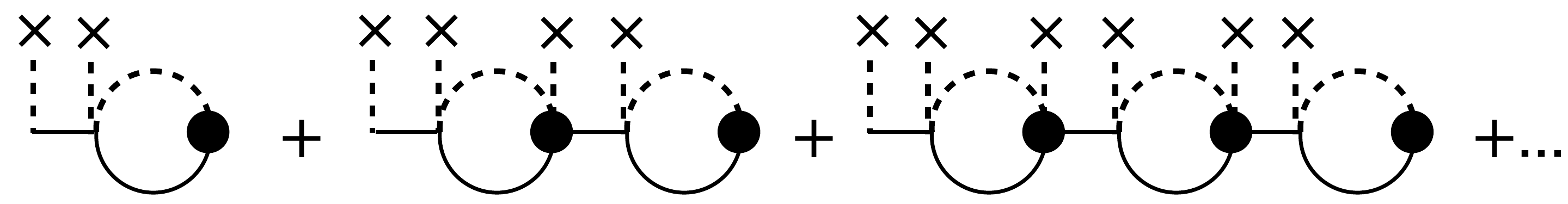} 
\caption{The diagrams containing the mixing between between the RPA diagrams ($\GRPA$) and $\GMFpole$ contributing to $G_3$.
There is another family of diagrams where the RPA vertex (the black dot) is attached to the left. This gives an identical contribution, hence the factor 2 in Eq.~(\ref{eq:three-1PR}). 
} 
\label{fig:1PR-three} 
\end{center} 
\end{figure}

The remaining part of the total Goldstino propagator is $G_3(p)$.
It is given by the diagrams whose two ends are connected with one condensate and one boson with finite momentum, and which are displayed in  Fig.~\ref{fig:1PR-three}. 
The resulting expression of $G_3(p)$ reads
\begin{align}
 \label{eq:three-1PR}
G_3(p)
&= -2U  \GMFpole (p) \tilde{G}(p),
\end{align}
which reduces to 
\begin{align}
 \label{eq:three-1PR-p=0}
G_3(\omega,\vzero)
&= -2\frac{\rhof\rhob}{\rho} 
\left(\frac{1}{\bar{\omega}}
-\frac{1}{\bar{\omega}+U\rho}\right),
\end{align}
at zero momentum.
The factor 2 reflects the fact that the condensate can be attached to either of the two ends of the diagrams.

The full Goldstino propagator is obtained by summing the three separate contributions that we have analyzed. 
By adding Eqs.~(\ref{eq:Gtil-1PR-p=0}), (\ref{eq:S-1PR-p=0}), and (\ref{eq:three-1PR-p=0}), one observes that the poles at $\bar\omega=-U\rho$ cancel among themselves, leaving, as expected, a single pole at $\bar\omega=0$ with a residue equal to the density,  in complete agreement with Eq.~(\ref{eq:G0-p=0-pole}).
The final result is the same as that obtained  without BEC~\cite{Blaizot:2015wba}, but how it emerges is completely different.

\subsubsection{Finite momentum case}
\label{ssc:finite-p}

We proceed now to the finite momentum case. 
First, we consider the energy/momentum region near the Goldstino pole, where we can safely expand $\GRPA$ and $\GMFpole$ as we did in Eq.~(\ref{Gexpanded}).
As a result, we obtain
\begin{align}
\label{eq:1PR-expansion-Gtil}
\begin{split}
{G}(p) 
&\simeq -\frac{Z}{\bar{\omega}-\alpha  \vp^2/(2m)}  
\end{split}
\end{align}
with  
\begin{align}
\label{eq:residue-total}
Z&=\rho
-\frac{4}{5}\rhof\left(\frac{|\vp|}{\kf}\frac{\eF}{U\rho}\right)^2 ,\\
\nonumber
\alpha&\equiv \frac{\rhob-\rhof}{\rho}+\frac{4}{5}\frac{\rhof}{\rho} \frac{\varepsilon_F}{U\rho}\\
\label{eq:alpha}
&= \alpha_s+\frac{4}{5}\frac{\rhof}{\rho} \frac{\varepsilon_F}{U\rho}.
\end{align}
These formulae reduce to Eqs.~(\ref{eq:Z-Gtilde}) and (\ref{eq:alphaf}) [(\ref{eq:Gtil-MF2})] when $\rhob=0$ [$\rhof=0$], as they should.
Also that the expression for $\alpha$ is the same as  that obtained in the absence of BEC~\cite{Blaizot:2015wba}.

\begin{figure}[t] 
\begin{center}
\includegraphics[width=0.48\textwidth]{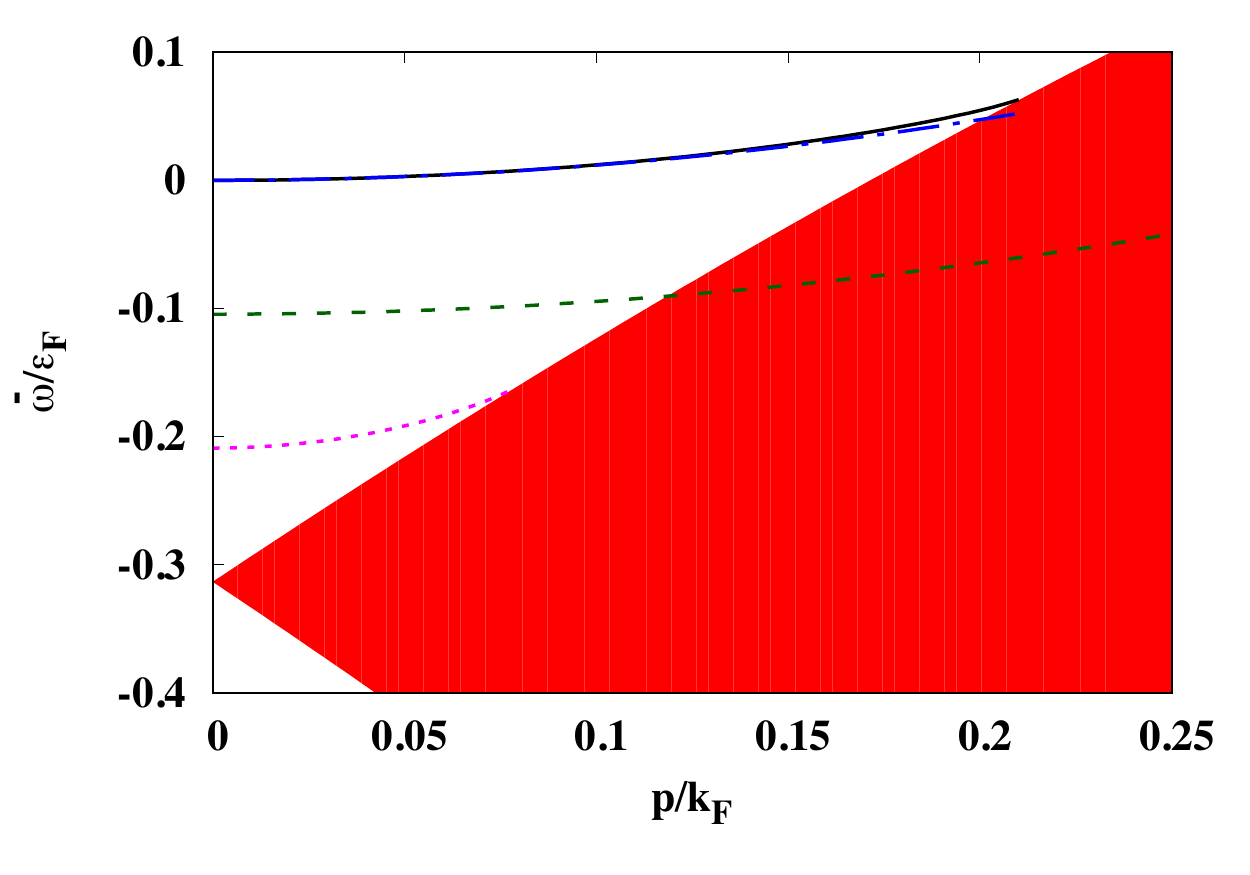}  
\caption{The range of the continuum (red shed area), the numerical result for the pole position of the Goldstino propagator
 (black solid line), and the pole position obtained from the small $|\vp|$ expansion, Eq.~(\ref{eq:alpha}) (blue long-dashed line).
For illustration of the ``level repulsion'', the pole positions of 
 $\GMFpole$ (green dashed line) and  $\GRPA$ (magenta dotted line) are also plotted. Note that at $\vp=\vzero$ the tip of the continuum corresponds to the fictitious pole at $\bar\omega=-U\rho$, carrying no spectral weight.  
The densities are the same as in Sec.~\ref{ssc:U=0}, i.e., $\rhob=2\rhof$, and the interaction strength is $U\rhof/\eF=0.1$.
} 
\label{fig:dispersion-smallp} 
\end{center} 
\end{figure}

The location of the Goldstino pole obtained numerically is plotted in  Fig.~\ref{fig:dispersion-smallp}, and compared to the approximate expression $\bar{\omega}=\alpha\vp^2/(2m)$. 
The interaction strength is set to  a small value, $U\rhof/\eF=0.1$, or $\kf a=0.3\pi/4\simeq 0.24 $ in terms of $a$,  for which the weak-coupling analysis is reliable. 
One sees on Fig.~\ref{fig:dispersion-smallp} that the approximate expression is accurate as long as $|\vp|\lesssim 0.16\kf$. This is indeed the expected range of validity of the expansion, namely $U\rho\gg \kf|\vp|/m$, as can be seen from the denominator of the expression of $\GMFcont$, Eq.~(\ref{eq:Gtil-MF0cont}).
This condition leads to $|\vp|\ll U\rho m/\kf\simeq 0.15\kf$ for the current values of the parameters.
Note that because  the continuum is shifted down by the MF correction $U\rho$, as compared to the free case (\ref{eq:continuum-range-U=0}),  the Goldstino pole remains out of the continuum as long as $|\vp|$ is smaller than $\sim 0.21\kf$.

Also plotted in Fig.~\ref{fig:dispersion-smallp} are the dispersion relations corresponding to  the poles of $\GMFpole$ (Eq.~(\ref{eq:Gtil-MF0pole2})) and $\GRPA$   (Eq.~(\ref{eq:Gtil-RPA})). 
This illustrates the effect of the level repulsion already discussed in the case $|\vp|=0$, yielding eventually the distribution of spectral weight between the continuum and the Goldstino pole. Of course, the supersymmetry plays a crucial role here in putting the Goldstino pole at $\bar{\omega}=0$ for $\vp=\vzero$.

\begin{figure}[t!]  
\begin{center}
\includegraphics[width=0.4\textwidth]{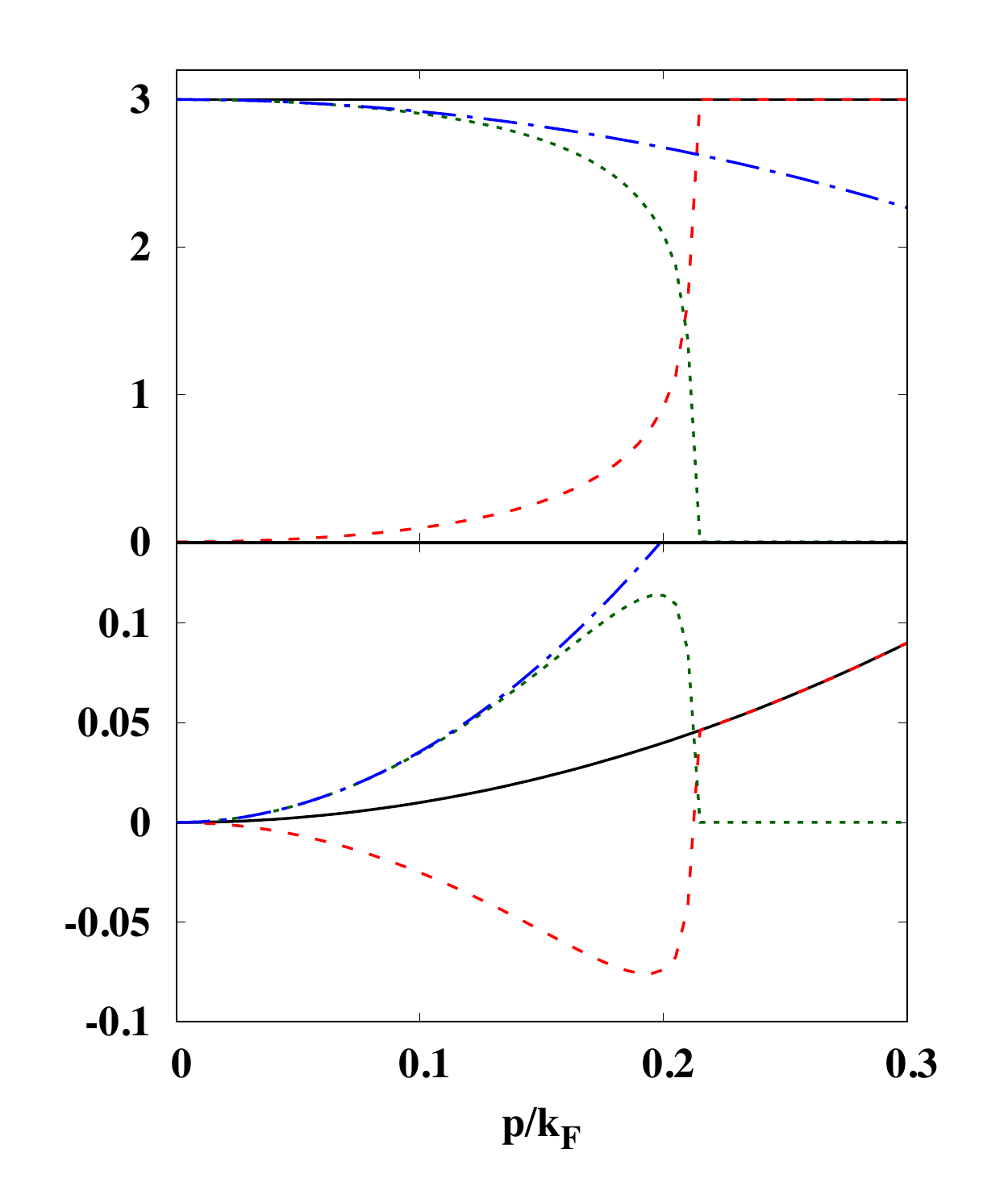}  
\caption{Upper panel: the contributions to the zeroth moment of $\sigma$ from the pole (green dotted line), the continuum (red dashed line), and their sum (black solid line). 
The pole contribution in the small momentum/energy expansion (blue long-dashed line) is also plotted.
Lower panel: the contributions to the first moment of $\sigma$ from the sources listed above.
The unit of energy is $\eF$  and that of $\sigma$ is $\rhof/\eF$. Densities and coupling are the same as in Fig.~\ref{fig:dispersion-smallp}.
} 
\label{fig:sumrule-total} 
\end{center} 
\end{figure}


\begin{figure}[t] 
\begin{center}
\includegraphics[width=0.48\textwidth]{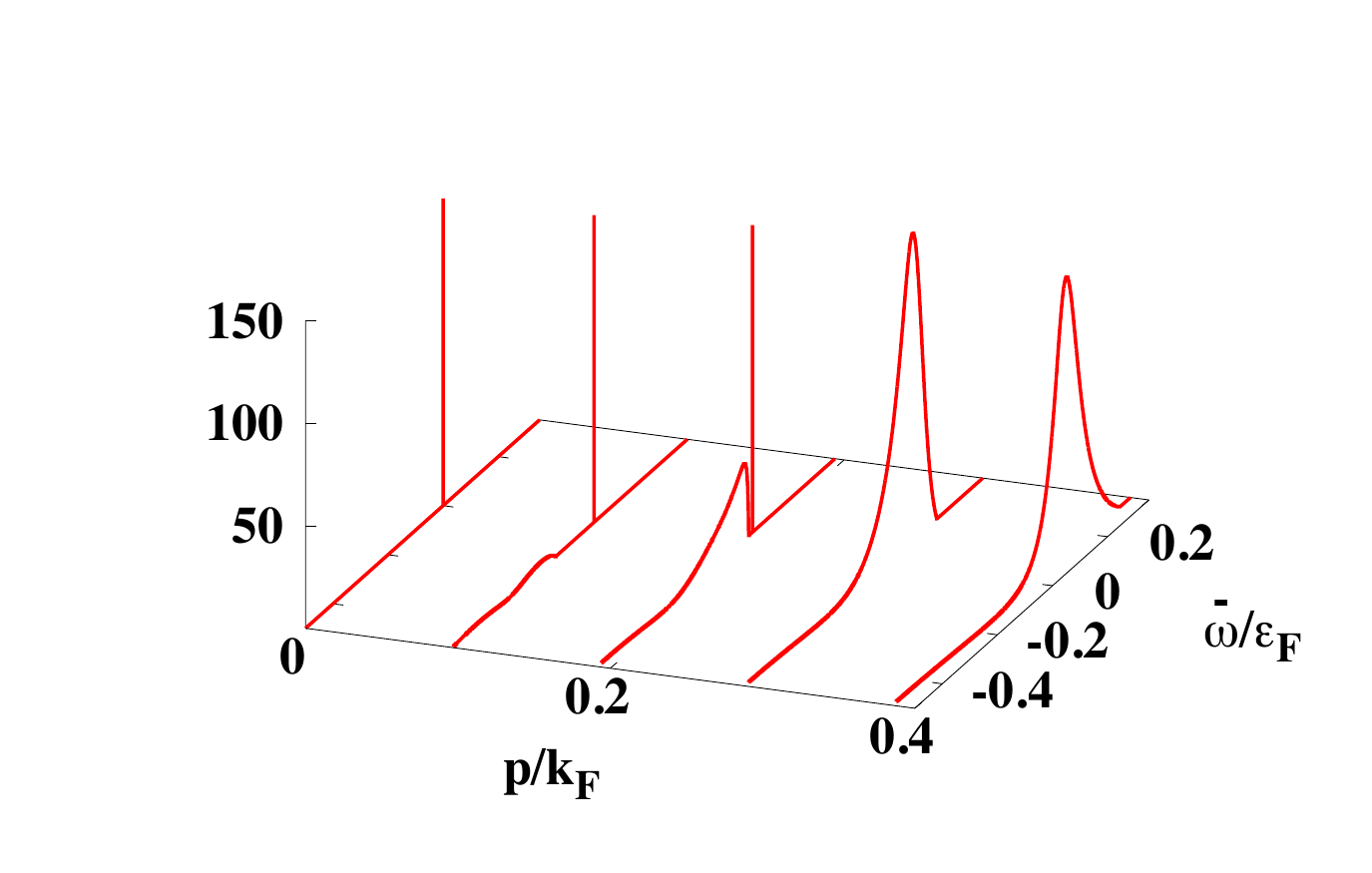} 
\caption{The spectral function of the Goldstino (in units of $\rhof/\eF$). Densities and coupling are the same as in Fig.~\ref{fig:dispersion-smallp}. At very small $\vp$ the continuum carries no spectral weight, this being entirely taken by the Goldstino pole. As the momentum increases, a peak develops in the continuum, eventually merging with the pole, leading to a broad peak whose width decreases with increasing momentum. 
} 
\label{fig:Goldstino-spectrum} 
\end{center} 
\end{figure} 

The spectral function is analyzed in more detail in Fig.~\ref{fig:sumrule-total}.
The contributions to the zeroth moment of $\sigma$ from the pole and the continuum are displayed in  the upper panel of this figure. At small momenta, $|\vp|\lesssim 0.11\kf$, these are well accounted for by the expansion (\ref{eq:residue-total}).
In the same plot, we see that the continuum contribution is suppressed for small momentum, with all the spectral weight being carried there by the Goldstino. 
The lower panel of Fig.~\ref{fig:sumrule-total} reveals large cancellation between the pole and continuum contributions to the first moment of the spectral function.
This can be understood as follows:
At small momentum, the pole contribution is found to be $\alpha\rho\vp^2/(2m)$ by using Eq.~(\ref{eq:residue-total}).
On the other hand, the sum rule (\ref{eq:sumrule2-G}) requires  the sum of the pole and the continuum contributions to be $\alpha_s\rho\vp^2/(2m)$, so the continuum contribution should be $(\alpha_s-\alpha)\rho\vp^2/(2m)$.
At weak coupling, the second term in Eq.~(\ref{eq:alpha}) dominates over the first term, i.e.,  $\alpha\gg \alpha_s$.
Thus, the pole (continuum) contribution is approximately $\alpha\rho\vp^2/(2m)$ ($-\alpha\rho\vp^2/(2m)$).
This behavior is the same as for the case without BEC~\cite{Blaizot:2015wba}.

When $|\vp|$ becomes larger than $\sim 0.21\kf$, the pole contribution vanishes since the pole merges with the continuum (see Fig.~\ref{fig:dispersion-smallp}).
The sum of the pole and the continuum contributions to the zeroth moment equals $\rho$, as it should because of the sum rule (\ref{eq:sumrule1-G}).
It implies that the spectral weight of the continuum increases rapidly around the momentum at which the pole is absorbed, which is demonstrated in Fig.~\ref{fig:sumrule-total}. 
These behaviors, namely the suppression (enhancement) of the continuum at small $|\vp|$ (above $|\vp| \simeq 0.21\kf $), can be seen also from the spectral function $\sigma$ plotted in Fig.~\ref{fig:Goldstino-spectrum}.

\section{Phenomenological implication}
\label{sec:phenomenology}

The strong coupling between the fermion and the Goldstino may offer a possibility to infer the properties of the Goldstino from the study of the fermion propagator. This is what we explore in this section. 
\begin{figure}[t] 
\begin{center}
\includegraphics[width=0.2\textwidth]{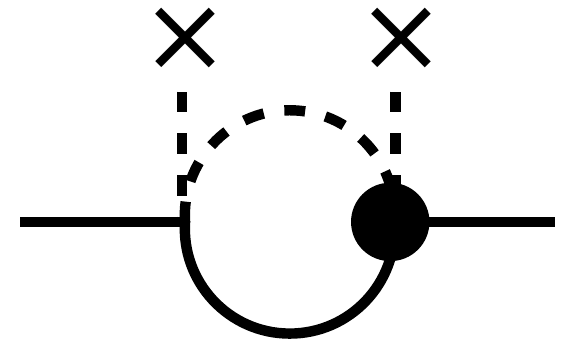} 
\caption{The fermion self-energy.
} 
\label{fig:fermionselfenergy} 
\end{center} 
\end{figure} 

 \begin{figure}[t] 
\begin{center}
\includegraphics[width=0.2\textwidth]{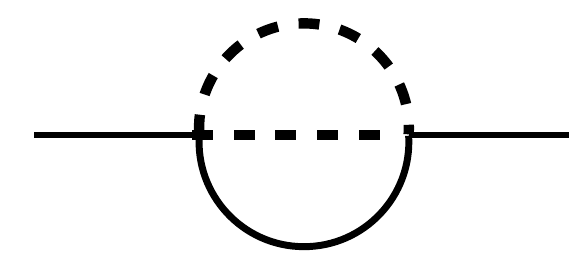} 
\caption{The fermion self-energy at the two-loop order, which we do not take into account.
} 
\label{fig:sigma-nonpinch} 
\end{center} 
\end{figure}

\subsection{Fermion spectrum}

As was mentioned in the Introduction, the experimental observation of the fermion spectrum at small momentum by using photoemission spectroscopy technique~\cite{Stewart: 2008} seems to be the most direct way to investigate the Goldstino spectrum. 
Actually, one of the component of the Goldstino propagator introduced in the previous section, $G_S$, is proportional to the fermion propagator:
\begin{align}
\label{eq:GS-SR}
G_S(p)
&= \rhob S(p).
\end{align}
This is reflected in the fact that all the diagrams in Fig.~\ref{fig:1PR-S} have this particular structure. The corresponding  fermion self-energy  is given by the diagram in Fig.~\ref{fig:fermionselfenergy}.
Note that the two-loop self-energy drawn in Fig.~\ref{fig:sigma-nonpinch} is not taken into account here,  although formally it is of the same order  as the one in Fig.~\ref{fig:fermionselfenergy}, namely ${\cal O}(U^2)$.
This is because in the small $\bar{\omega}$ and $|\vp|$ region, the contribution from the diagram in Fig.~\ref{fig:fermionselfenergy} is enhanced, as we have seen in the previous section, while that from Fig.~\ref{fig:sigma-nonpinch} is not.\footnote{In two dimensions, the contribution of this diagram has an infrared divergence. Although this was not considered in \cite{Blaizot:2015wba}, and we have not carried out a detailed analysis, it is possible that this leads to an enhanced contribution of the corresponding process in two dimensions,  similar to that of a BEC in the three-dimensional set up considered in the present paper.}

The fermion retarded propagator and its spectral function $\sigma_S$ can be deduced from Eq.~(\ref{eq:S-1PR}). 
By doing the small $\bar{\omega}$ and $\vp$ expansion, we get 
\begin{align}
{S}(p) 
&\simeq -\frac{Z_S}{\bar{\omega}-\alpha\vp^2/(2m)}, 
\end{align}
where 
\begin{align}
\label{eq:residue-S}
Z_S&\equiv 
 \frac{\rhob}{\rho}
 \left[1 -\frac{\vp^2}{\kf^2}\frac{4\varepsilon_F\rhof}{U\rho^2}\left(-1+\frac{3\varepsilon_F}{5U\rho}\right)\right] .
\end{align}
We see that the dispersion relation is the same in the total Goldstino propagator, 
which is natural because of the strong mixing discussed in the previous section.
This property suggests that, for experimental investigation of the dispersion relation of the Goldstino, it is sufficient to focus on detecting the fermion spectrum.
The presence of BEC is important here: in the absence of BEC, the  mixing process between ${G}_{\text{RPA}}$ and $G^{\mbox{\tiny MF}}_{\rm pole}$, which yields the self-energy in Fig.~\ref{fig:fermionselfenergy}, disappears.

\begin{figure}[t!]  
\begin{center}
\includegraphics[width=0.5\textwidth]{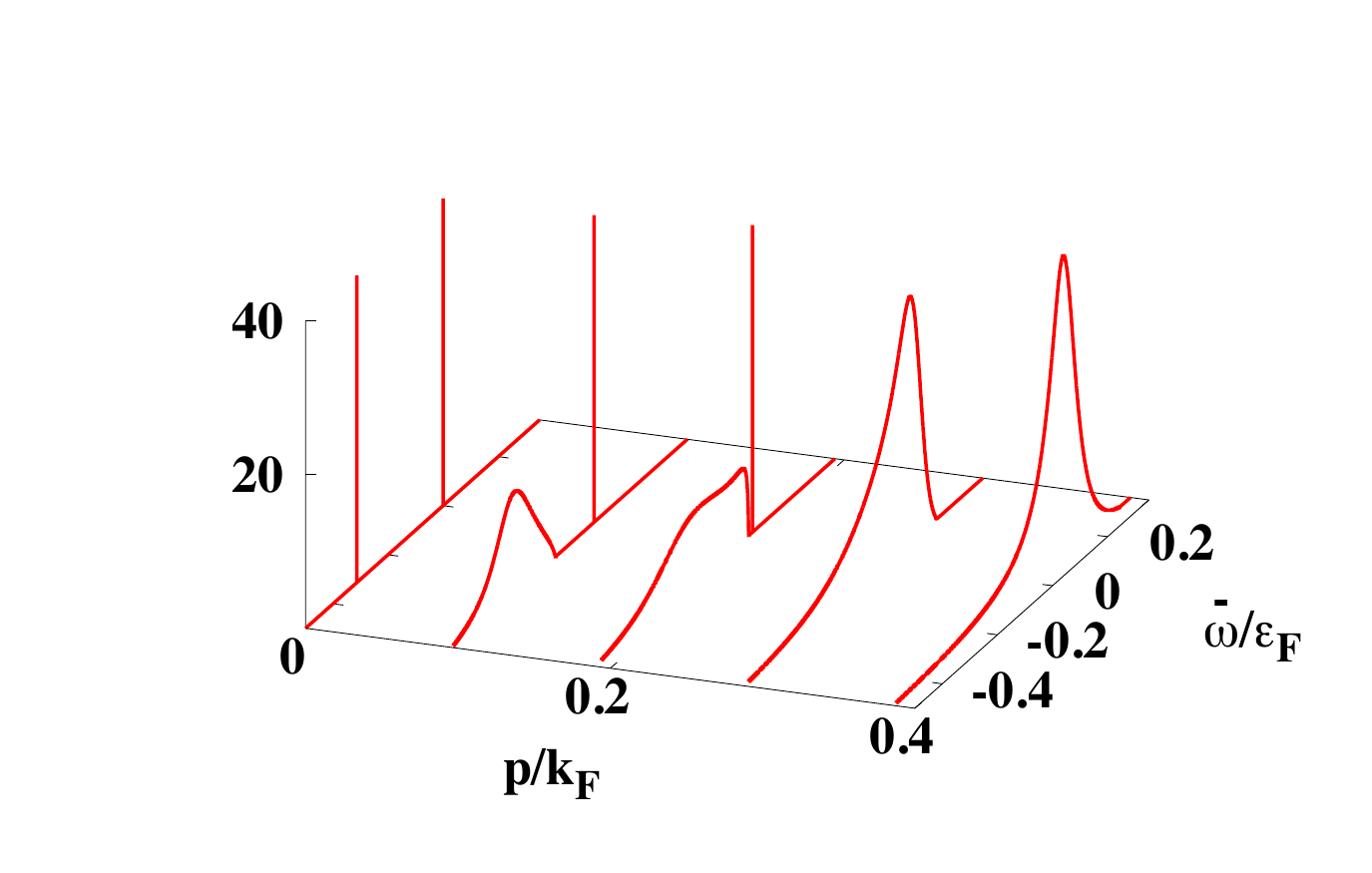}  
\caption{The fermion spectral function $\sigma_S$ as a function of $|\vp|$ and $\bar{\omega}$.
The unit of $\sigma_S$ is $1/\eF$. Densities and coupling are the same as in Fig.~\ref{fig:dispersion-smallp}. At $\vp=\vzero$, the continuum ends in a pole with spectral weight $\rho_f/\rho$, while the spectral weight of the other pole is $\rho_b/\rho$.
} 
\label{fig:spectrum} 
\end{center} 
\end{figure}

\begin{figure}[t!]  
\begin{center}
\includegraphics[width=0.45\textwidth]{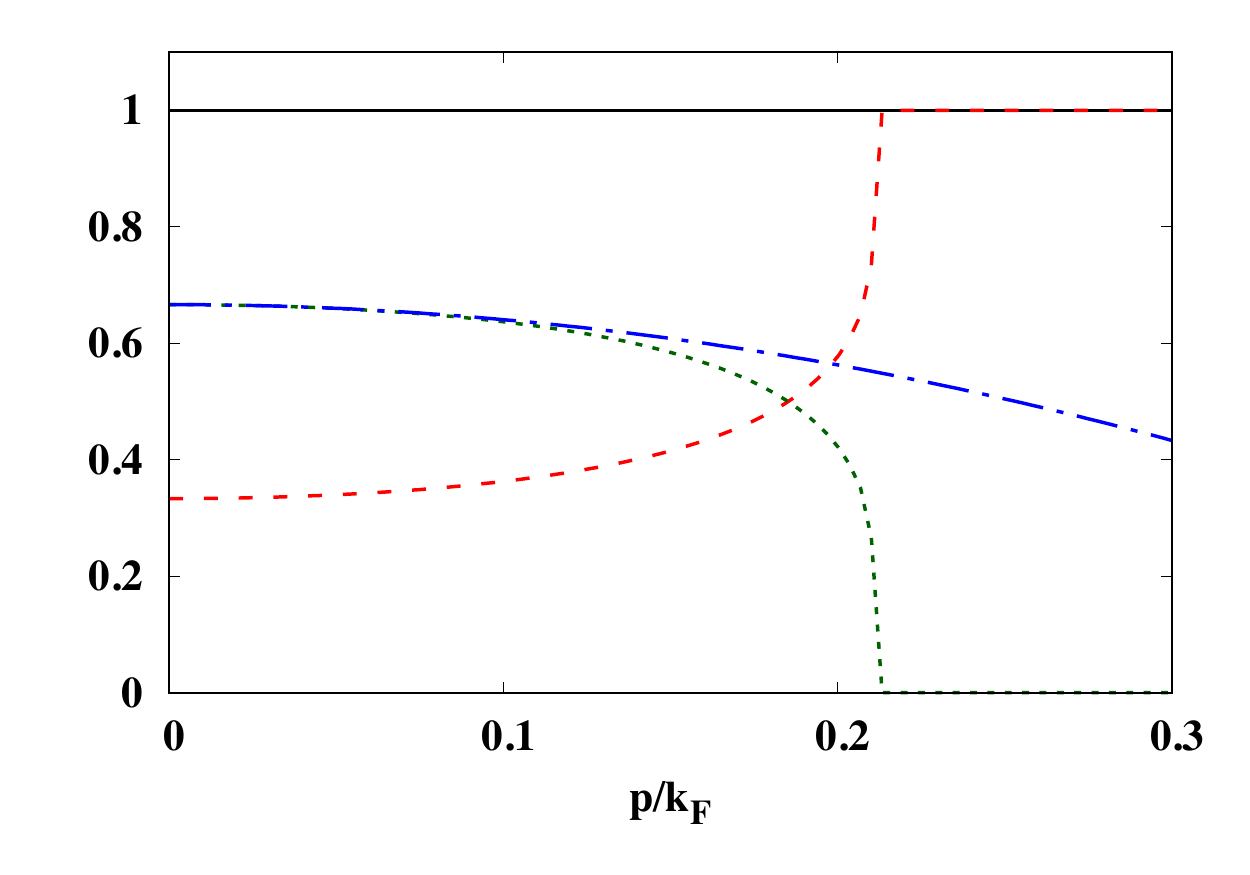}  
\caption{Continuum and pole contributions to the zeroth moment of $\sigma_S$.
The contributions  from the pole (green dotted line), the continuum (red dashed line), and their sum (black solid line) are plotted. 
The pole contribution in the small momentum/energy expansion (blue long-dashed line) is also plotted for comparison. Densities and coupling are the same as in Fig.~\ref{fig:dispersion-smallp}. Thus, the respective spectral weights of the pole and the continuum at $\vp=\vzero$ are respectively $\rho_b/\rho=2/3$ and $\rho_f/\rho=1/3$.
} 
\label{fig:sumrule-S} 
\end{center} 
\end{figure}

In addition to this Goldstino pole, the fermion spectral function contains also a continuum, as can be seen from the plot for $\sigma_S$ in Fig.~\ref{fig:spectrum}.
For $|\vp|$ smaller than $\sim 0.21\kf$,  the fermion spectral function is   well described by\begin{align}
\label{eq:spectral-S-cont-pole}
\sigma_S(p)
&= 2\pi Z_S \delta \left(\bar{\omega}-\alpha \frac{\vp^2}{2m}  \right)
+\theta(\kf-\kcf)\sigma^{{\text{cont}}}_S(p).
\end{align}
Here $\kcf\equiv m|\bar{\omega}+\vp^2/(2m)+U\rho|/|\vp|$, which is shifted by the MF effect compared with the free case.
We do not write the expression for $\sigma^{{\text{cont}}}_S(p)$ since it is not relevant to our discussion.
We see that the continuum and the pole have comparable spectral weights, as can be seen from Fig.~\ref{fig:sumrule-S}.
This is quite different compared with the total Goldstino propagator that we discussed in the previous section, in which the continuum is suppressed for small $|\vp|$. Here the continuum ends at $\vp=\vzero$ in a pole which carries a fraction $\rho_f/\rho$ of the spectral weight. The ``regular'' pole carries a fraction $\rho_b/\rho$, as can be deduced from Eq.~(\ref{eq:residue-S}).
The total spectral weight is equal to unity, in agreement with the well-known sum rule,
\begin{align}
\label{eq:sumrule1-S}
\int \frac{d\bar{\omega}}{2\pi}\sigma_S(p)
&= 1.
\end{align}
When the momentum exceeds $\sim 0.21\kf$, the pole merges with the continuum, and the whole spectral weight is then carried by the continuum.
The width of the peak in the continuum is decreasing function of $|\vp|$ for $|\vp|\gtrsim 0.3\kf$. 
This is to be expected since, when $|\vp|$ becomes large, the interaction becomes negligible in Eq.~(\ref{eq:S-1PR}), and the fermion spectral function  approaches the free value, 
\begin{align}
\sigma_S(p)
&= 2\pi  \delta \left(\bar{\omega}-\frac{\vp^2}{2m}\right),
\end{align}
whose width is zero.
When the momentum exceeds $0.85\kf$, the peak leaves the continuum and it becomes a pole with zero width.
One can see this point in the upper-right panel of Fig.~\ref{fig:distribution}.
Thus we have seen the crossover of the fermion spectrum from small $\vp$ (Goldstino pole and the continuum) to large $\vp$ (pole in the free limit).

\begin{figure}[t] 
\begin{center}
\includegraphics[width=0.48\textwidth]{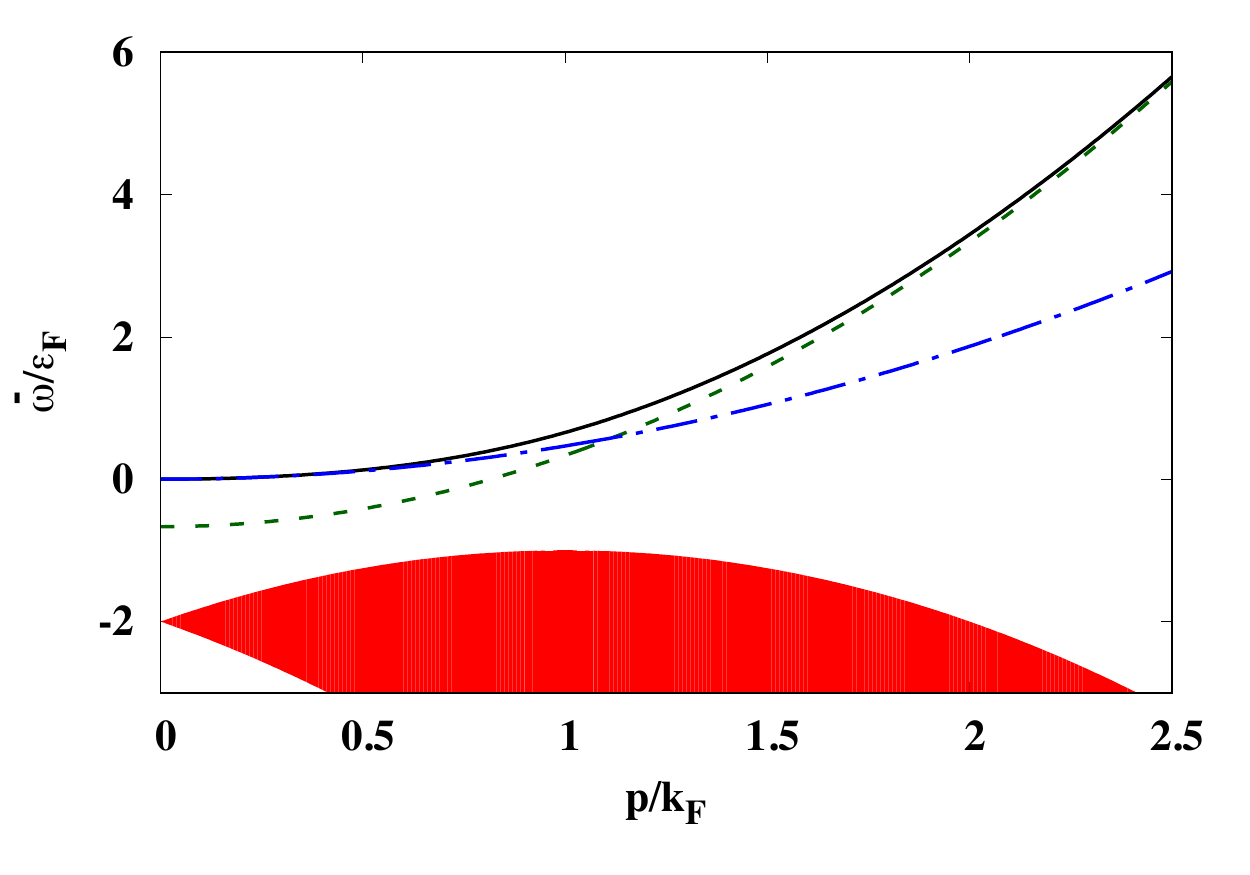}  
\caption{The range of the continuum (red shed area), the  pole position (black solid line), and the pole position of $G^{\mbox{\tiny MF}}_{\rm pole}$ (green dashed line) are plotted. The blue long dashed line represents the small momentum approximation to the dispersion relation. 
The densities are the same as in Fig.~\ref{fig:dispersion-smallp}, and $U=U_{c1}$. 
} 
\label{fig:dispersion-middlelargeU} 
\end{center} 
\end{figure}

\begin{figure}[t!]  
\begin{center}
\includegraphics[width=0.45\textwidth]{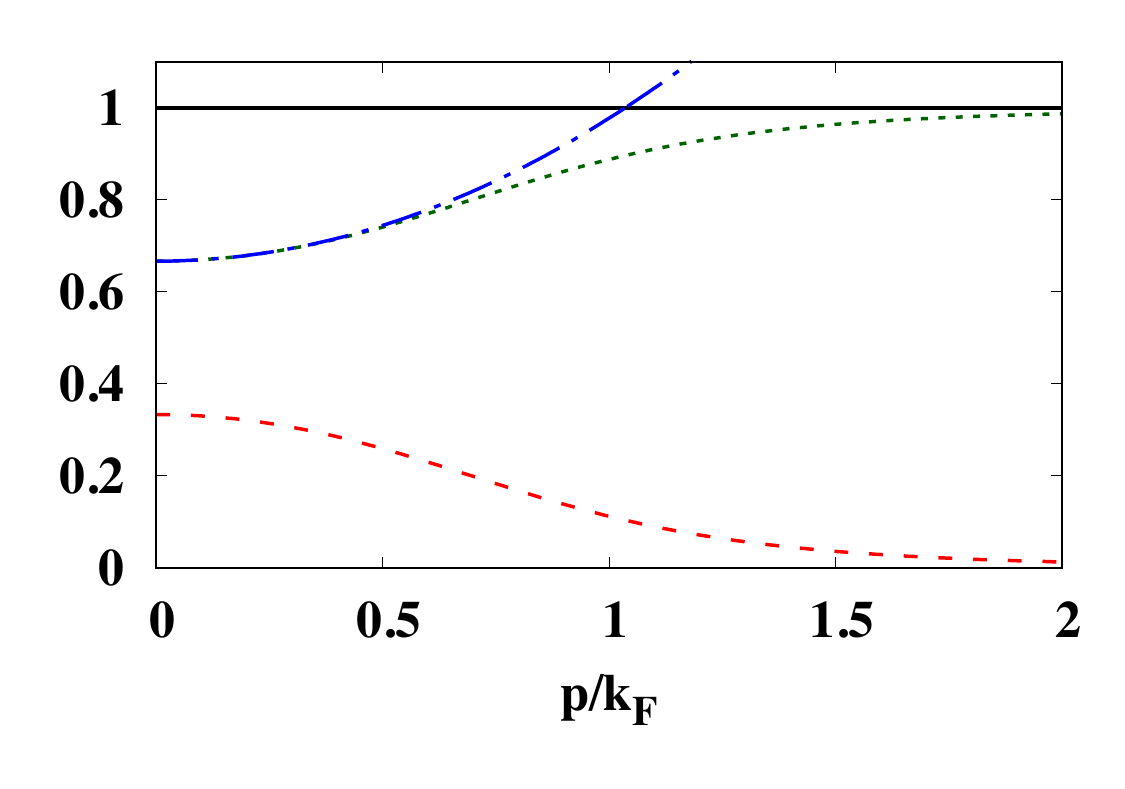}  
\caption{The contributions from the continuum (red dashed line) and the pole (green dotted line) to the zeroth moment of $\sigma_S$, and their sum (black solid line) at $U=U_{c1}$.
The pole contribution in the small momentum expansion (blue long-dashed line) is also plotted.  The densities are the same as in Fig.~\ref{fig:dispersion-smallp} ($\rho_b=2\rho_f$). Thus, the respective spectral weights of the pole and the continuum at $\vp=\vzero$ are respectively $\rho_b/\rho=2/3$ and $\rho_f/\rho=1/3$.
} 
\label{fig:sumrule-S-middlelargeU} 
\end{center} 
\end{figure}

Finally, let us  see how the fermion spectrum is modified when we increase the interaction strength to the maximum allowed value $U=U_{c1}$ (see Eq.~(\ref{eq:stability-condition})).
As can be seen in Fig.~\ref{fig:dispersion-middlelargeU},  the Goldstino pole now always lies outside  the continuum  because the gap between the pole and the continuum at $|\vp|=0$, which is equal to $U\rho$, is large enough.
We also see that the position of the pole is approximately equal to the value obtained in the small momentum expansion ($\bar{\omega}=\alpha\vp^2/(2m)$) for quite a large range of momenta, $|\vp|\lesssim 0.7\kf$, and it approaches the MF value ($\bar{\omega}=\vp^2/(2m)-U\rhof$) for $|\vp|\gtrsim 1.9 \kf$.
We also plot the contributions from the pole and the continuum to the zeroth moment for $\sigma_S$ in Fig.~\ref{fig:sumrule-S-middlelargeU}.
The pole contribution agrees with the value in the small momentum expansion (\ref{eq:residue-S}) quite well at small momentum.
Because the pole is never absorbed by the continuum, its contribution does not decrease as a function of $|\vp|$ but increases, in contrast to the weak coupling case.
At large momentum, it carries the total spectral weight.


\subsection{Fermion distribution}

\begin{figure*}[t] 
\begin{center}
\includegraphics[width=0.7\textwidth]{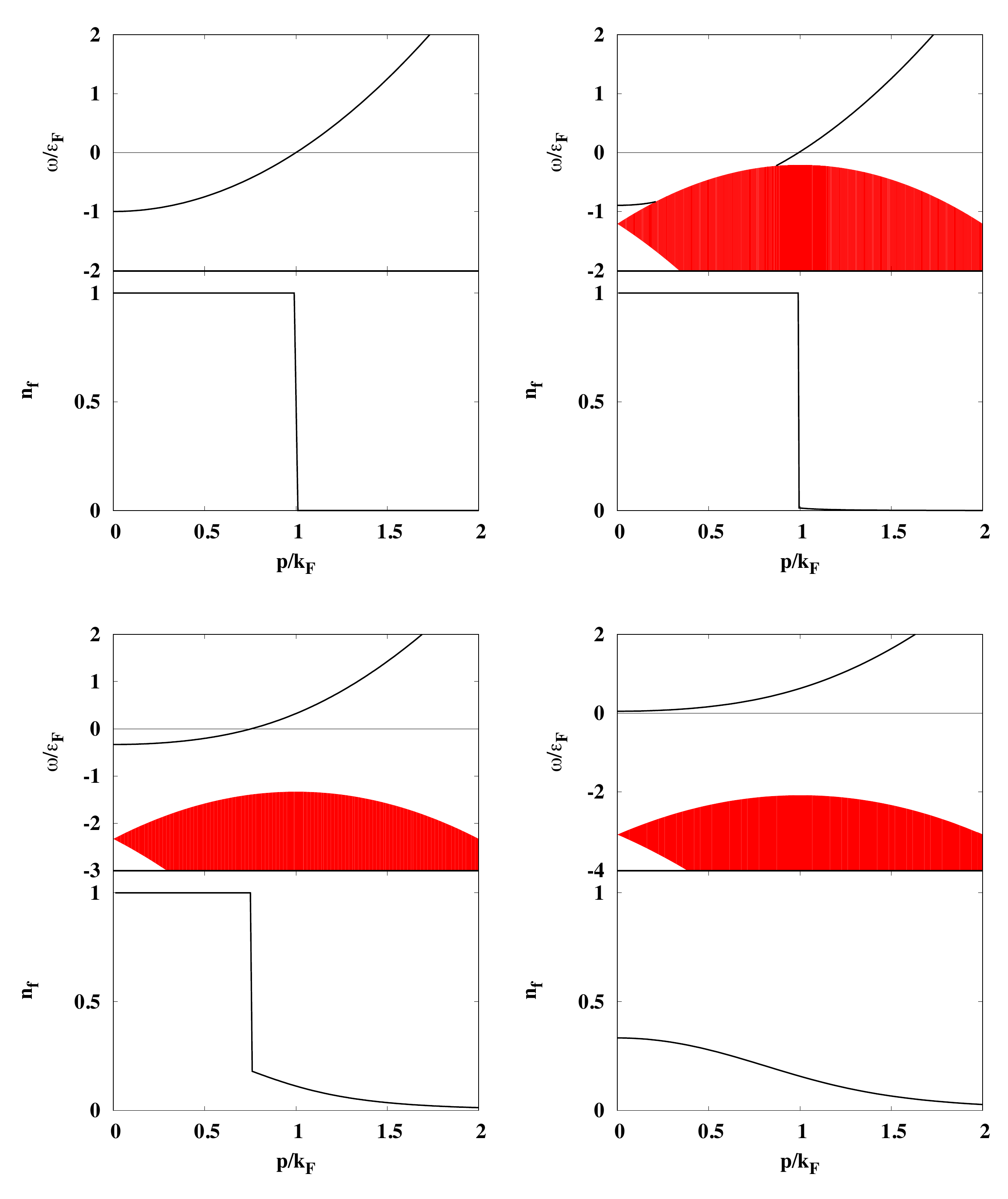} 
\caption{Upper parts of the panels: The  continuum region (red shaded area) and the pole position (black solid curve).
The line $\omega=0$ is also plotted.
Lower parts of the panels: Fermion distribution in momentum space.
The upper-left panels describe the free limit ($U=0$).
The remaining panels, upper-right, lower-left, and lower-right ones, correspond to the case of $U\rhof/\eF=0.10$, $U=U_{c1}$, $U=U_{c2}$, respectively.
} 
\label{fig:distribution} 
\end{center} 
\end{figure*}

The other quantity that may have the possibility of experimental observation is the fermion momentum distribution.
This is related to the spectral function by
\begin{align}
\label{eq:nf-spectrum}
\nf(\vp)
&= \int^\infty_{-\infty} \frac{d\omega}{2\pi}
\nf(\omega)\sigma_S(\omega,\vp),
\end{align}
where $\nf(\omega)=\theta(-\omega)$.
In the free limit, where the spectral function is given by $\sigma_S(\omega,\vp)= 2\pi \delta(\omega-\epsilon^f_{\vp})$ with $\epsilon^f_{\vp} = \epsilon^0_\vp-\muf$, the distribution $\nf(\vp)$ reduces to
\begin{align}
\label{eq:nf-free}
\nf(\vp)
&=\theta(\kf-|\vp|).
\end{align}
Such a situation is visualized in the upper-left panels of Fig.~\ref{fig:distribution}.
Note that the pole position is plotted there in terms of $\omega$ instead of $\bar{\omega}$, in order to show explicitly the physical excitation energy without the constant shift coming from the chemical potential difference. 

When we turn on the interaction, the fermion spectrum is modified. This is illustrated in Fig.~\ref{fig:distribution}. When the interaction is weak ($U\rhof/\eF=0.1$), there is not much modification of the momentum distribution. This is because the entire spectral weight resides mostly at negative $\omega$. The contribution of the Goldstino pole crosses the line $\omega=0$ for a momentum nearly equal to $k_F$, and stops to contribute for larger momenta. A tiny contribution to the momentum distribution is visible just above $k_F$.

In order to see more clearly the modification of the fermion momentum distribution induced by the coupling to the Goldstino, we increase the coupling up to $U=U_{c1}$. The results are displayed in the lower-left in Fig.~\ref{fig:distribution}. 
The Fermi sea is largely distorted, so that the new Fermi surface is located at $|\vp| \sim 0.7\kf$.
Also, the momentum distribution extends above the Fermi surface, by a small contribution which decreases with $|\vp|$.
To understand these results, recall that the spectral function at small momentum is  well described by the Goldstino pole and the continuum, Eq.~(\ref{eq:spectral-S-cont-pole}).
Equation~(\ref{eq:nf-spectrum}) yields then
\begin{align}
\label{eq:nf-smallp}
\nf(\vp)
&= Z_S \theta\left(\Delta\mu-\alpha\frac{\vp^2}{2m}\right)
+(1-Z_S),
\end{align}
where the first term comes from the Goldstino pole and the second one from the continuum.
Here we have used the fact that the continuum is always in the negative energy region (maximum value: $\omega=-U\rhob$ at $|\vp|=k_F$), and the property
\begin{align}
\label{eq:sumrule-continuum}
\int^\infty_{-\infty} \frac{d\omega}{2\pi}\theta(k_F-k_{cf})\sigma_{\text{cont}}(p)
&= 1-Z_S,
\end{align}
which follows from the sum rule (\ref{eq:sumrule1-S}).
As for the Goldstino pole contribution, we note that this vanishes when the Goldstino pole sits at positive $\omega$, which occurs  when $|\vp|>k_\alpha\equiv\sqrt{2m \Delta\mu/\alpha}$. When $|\vp|<k_\alpha$, the entire spectral weight resides at negative $\omega$ and contributes unity to the momentum distribution function. When $|\vp|>k_\alpha$, the Goldstino pole stops to contribute to the momentum distribution, only the continuum does, by an amount equal to $1-Z_S$. As the momentum increases, so does $Z_S$. Eventually at large enough momentum, the Goldstino carries the entire spectral weight and the momentum distribution vanishes.

As  the interaction strength grows, $k_\alpha$ decreases, and eventually vanishes. This occurs when
$\Delta\mu=\varepsilon_F-U\rho_f$ changes sign from positive to negative, i.e.,  when $U$ exceeds the critical value $U_{c2}\equiv \varepsilon_F/\rhof$. Strictly speaking, since $U_{c2}>U_{c1}$ this value of the coupling constant is outside the region allowed by the condition (\ref{eq:stability-condition}). 
It is nevertheless interesting to explore what happens then, since  the fermion distribution function drastically changes, as can be seen in the lower-right panels of Fig.~\ref{fig:distribution}, in particular in the right panel where the Fermi surface has disappeared.  
As the interaction strength increases, moving from the left to the right panel of the lower part of Fig.~\ref{fig:distribution}, the discontinuity of the Fermi surface decreases. 
This discontinuity is given by the residue $Z_S$ evaluated at the new Fermi momentum (note that the momentum distribution stays equal to unity for momenta below this momentum). 
When the interaction strength reaches the value $U_{c2}$, the location of the Fermi surface (i.e, the singularity in the momentum distribution) has moved to $|\vp|=0$ (the fermion pole touches the $\omega=0$ line at $|\vp|=0$), and the residue there is $\rho_b/\rho=0.66$ [See Eq.~(\ref{eq:residue-S})] while it is $\simeq 0.8$ at $U=U_{c1}$ as can be seen from Fig.~\ref{fig:distribution}. 
Note that the discontinuity remains finite as its location reaches $|\vp|=0$. From that point on, as one continues to increase the coupling, the entire contribution to the momentum distribution comes from the continuum, and the momentum distribution is completely smooth.

\section{Conclusions and outlook}
\label{sec:summary}

We have analyzed the spectral properties of the Goldstino excitation in a supersymmetric mixture of Bose and Fermi cold atoms, with the bosons forming a BEC at zero temperature. At leading-order  in the weak coupling regime the excitations can  be studied within the RPA, taking into account the mixing processes between the supercharge and the fermion. The way the collective excitation develops, depending on the values of the various parameters characterizing the system, turned out to be in itself an interesting investigation in many-body physics. However, it would be even more interesting if such excitations could be observed in appropriate experimental setups. In this perspective, we have noted that 
 the mixing between the fermion and the Goldstino produces a strong modification of the  fermion spectral function.
This  could be reflected in experimental observables, such as the fermion spectrum and the fermion momentum distribution.

Our analysis  is well-founded at small coupling but, as we have discussed,  the possible interesting effects on the fermion properties manifest themselves more visibly when the coupling gets stronger. It would  therefore be useful to consider the corrections to the present picture that arise when the coupling is pushed to the maximum  strength allowed by stability considerations. Among these corrections, an important one could be that arising from the phonons. Finite temperature effects, or damping mechanisms of the Goldstino are also worth investigating. 
We leave these interesting tasks for the future.

As a final remark, we comment on the possible realization of the supersymmetric setup for the Bose-Fermi mixture.
As was mentioned in Sec.~\ref{sec:intro2}, having equal boson and fermion masses, and tuning the two interaction strengths equal, are necessary to realize SUSY.
Among the Bose-Fermi mixtures realized currently, a ${}^6$Li-${}^7$Li mixture~\cite{Ferrier-Barbut} may provides us with a chance to realize SUSY since tuning their interaction strengths is relatively easy, and the mass ratio, $7/6\simeq 1.17$, is not very different from unity.  
Another candidate is a ${}^{173}$Yb-${}^{174}$Yb mixture~\cite{Hara}.
Though tuning the interaction strength is not as easy as the other candidate, their masses are almost equal.

\section*{Acknowledgement}

We thank Sandro Stringari and Stefano Giorgini for fruitful discussions and their comments on possibility for detecting Goldstino spectrum in experiment.
We also thank Shimpei Endo for his comment on possible realization of supersymmetric Bose-Fermi mixture.
The research of D. S. is supported by Alexander von Humboldt Foundation. 
The research of Y. H. is partially supported by JSPS KAKENHI Grants Numbers 15H03652, 16K17716 and the interdisciplinary Theoretical and Mathematical Sciences Program (iTHEMS).
We thank the hospitality of European Centre for Theoretical Studies in Nuclear Physics and Related Areas (ECT*), at which part of this work was done.
J.-P. B. thanks P. Fayet for interesting exchanges on this subject.

\appendix*

\section{Contribution from phonons}
\label{app:phonon}

In this Appendix, we estimate the contribution of the low momentum bosonic excitations to the Goldstino retarded Green's function $\tilde{G}$, at the one-loop order. 
At small momentum the bosonic excitations are phonons, and are well described by Bogoliubov theory. 
The phonon operator ($\alpha_\vk,\alpha_\vk^\dagger $) are related to the original boson operators ($b_{\vk},b_{\vk}^\dagger$) by the Bogoliubov transformation
\begin{align}
b_{\vk}&=u_\vk \alpha_\vk +v^*_{-\vk}\alpha^\dagger_{-\vk} , \\
b^\dagger_{\vk}&= u^*_\vk \alpha^\dagger_\vk +v_{-\vk}\alpha_{-\vk},
\end{align}
with the coefficients $u_\vk$ and $v_{\vk}$ given by
\begin{align}
u^2_\vk&= \frac{1}{2}\left[1+\frac{(\vk^2/(2m)+U\rho_b)}{\epsilon_\vk}\right],\\
v^2_\vk&=\frac{1}{2}\left[-1+\frac{(\vk^2/(2m)+U\rho_b)}{\epsilon_\vk}\right].
\end{align}
The phonon dispersion relation reads
\begin{align}
\epsilon^2_\vk &= \frac{\vk^2}{2m}\left( \frac{\vk^2}{2m}+ 2U\rho_b\right) .
\end{align}
It becomes linear ($\epsilon_\vk\simeq c|\vk|$) for small momentum, where $c\equiv \sqrt{U\rhob/m}$. 
At large momentum, the interaction effect becomes negligible and the dispersion relation remains that of a free boson, $\epsilon_\vk\simeq \vk^2/(2m)$.
 The characteristic momentum at which the behavior of the spectrum changes from linear to quadratic is $k_c\equiv \sqrt{mU\rhob}$.

In terms of the phonon excitations, the one-loop propagator $\tilde{G}$ reads
\begin{align} 
\label{eq:Gtil-phonon}
\begin{split}
\tilde{G}(p)
&=- \int \frac{d^3\vk}{(2\pi)^3}
\Biggl[|u_{\vk-\vp}|^2\frac{\nf(\ef)}{\epsilon_{\vk-\vp}-\ef+\omega} \\
&~~~+|v_{\vk-\vp}|^2\frac{1-\nf(\ef)}{-\epsilon_{\vk-\vp}-\ef+\omega}
\Biggr],
\end{split}
\end{align}
where $\ef\equiv \epsilon^0_\vk-\muf +U\rhob $.
In the weak coupling limit, the characteristic momentum $k_c$ is much smaller than the Fermi momentum, $k_c\ll \kf $. Then the numerator in the second term makes the contribution of this term to the low momentum region $|\vk|<k_c$, negligible.
Consider for instance the case $|\vp|=0$, where the contribution of the soft modes ($k\le k_c$) to  Eq.~(\ref{eq:Gtil-phonon}) reduces to
\begin{align} 
\tilde{G}(\omega,\vzero)
&=- \frac{1}{2\pi^2}\int^{k_c}_{0} d|\vk| |\vk|^2 \frac{|u_{\vk}|^2}{\epsilon_{\vk}-\ef+\omega} .
\end{align}
In the integration region, $|u_{\vk}|^2$ is of order $U\rho/(c|\vk|)$ and the denominator of the integrand is of order $U\rho$ when $\bar{\omega}$ is small.
Combining these estimates, one finds that $\tilde{G}$ is of order $k^2_c/c \sim m^{3/2}\sqrt{U\rho}$.
This is much smaller than the contribution from the momentum region $|\vk|>k_c$, which is shown to be $1/U$ in the main text, as long as $U$ is small enough.
This is of course due to the small phase space volume occupied by the soft modes, which is of order $k^3_c$.

The contribution of the low momentum region to other quantities such as $G_S$ and $G_3$ can also be shown to be small in a similar way.


\end{document}